\documentclass[10pt,journal]{IEEEtran}

\usepackage{amssymb}
\usepackage{amsmath}
\usepackage{cite}
\usepackage{url}
\usepackage{xcolor}
\usepackage{cite,graphicx,amsmath,amssymb}
\usepackage{fancyhdr}
\usepackage{mdwmath}
\usepackage{mdwtab}
\usepackage{caption}
\usepackage{amsthm}
\usepackage{setspace}
\usepackage{hyperref}
\usepackage{algorithm}
\usepackage{algorithmic}
\usepackage{multirow}
\usepackage{makecell}
\usepackage{mathtools}
\usepackage{subcaption}
\usepackage{bm}
\usepackage{tikz}
\usepackage{xurl}
\usepackage{stfloats}
\usepackage{mathrsfs}

\usepackage{booktabs}
\usepackage{multirow}
\usepackage{siunitx}
\usepackage{balance}

\hypersetup{colorlinks=true,
linkcolor=blue,
citecolor=blue,      
urlcolor=black,
}

\graphicspath{{./src/img/}}

\DeclareMathOperator*{\argmin}{arg\,min}
\newtheorem{remark}{Remark}
\newtheorem{theorem}{Theorem}

\newtheorem{lemma}{Lemma}

\newtheorem{corollary}{Corollary}

\newtheorem{proposition}{Proposition}

\allowdisplaybreaks
\NewDocumentCommand{\multiubrace}{mmm}
 {
  \egreg_multiubrace:nnn {#1} {#2} {#3}
 }

\title{Beamforming Design for Continuous Aperture Array (CAPA)-Based MIMO Systems}

\author{
        Zhaolin Wang,~\IEEEmembership{Member, IEEE}, Chongjun Ouyang,~\IEEEmembership{Member, IEEE},\\ and Yuanwei Liu,~\IEEEmembership{Fellow, IEEE}
\thanks{Zhaolin Wang and Yuanwei Liu are with the Department of Electrical and Electronic Engineering, The University of Hong Kong, Hong Kong (e-mail: \{zhaolin.wang, yuanwei\}@hku.hk).}
\thanks{Chongjun Ouyang is with the School of Electronic Engineering and Computer Science, Queen Mary University of London, London E1 4NS, U.K. (e-mail: c.ouyang@qmul.ac.uk).} 
}

\begin{document}

\maketitle
\begin{abstract}
    

    An efficient beamforming design is proposed for continuous aperture array (CAPA)-based point-to-point multiple-input multiple-output (MIMO) systems. In contrast to conventional spatially discrete array (SPDA)-MIMO systems, whose optimal beamforming can be obtained using singular-value decomposition, CAPA-MIMO systems require solving the eigendecomposition of a Hermitian kernel operator, which is computationally prohibitive. 
    To address this challenge, an explicit closed-form expression for the achievable rate of CAPA-MIMO systems is first derived as a function of the continuous transmit beamformer. Subsequently, an iterative weighted minimum mean-squared error (WMMSE) algorithm is proposed, directly addressing the CAPA-MIMO beamforming optimization without discretization approximation. Closed-form updates for each iteration of the WMMSE algorithm are derived via the calculus of variations (CoV) method. For low-complexity implementation, an equivalent matrix-based iterative solution is introduced using Gauss-Legendre quadrature.
    Our numerical results demonstrate that 1) CAPA-MIMO achieves substantial performance gain over the SPDA-MIMO, 2) the proposed WMMSE algorithm enhances performance while significantly reducing computational complexity compared to state-of-the-art Fourier-based approaches, and 3) the proposed WMMSE algorithm enables practical realization of parallel, non-interfering transmissions.
\end{abstract}

\begin{IEEEkeywords}
    Beamforming optimization, continuous aperture array (CAPA), MIMO 
\end{IEEEkeywords}

\section{Introduction} \label{sec:intro}


\IEEEPARstart{I}N the evolving landscape of multiple-input multiple-output (MIMO) communications, the continuous aperture array (CAPA) emerges as a transformative technology, redefining the paradigms of antenna design and performance \cite{comsoc_news, bjornson2019massive, liu2024capa}. \textcolor{black}{Unlike traditional spatially discrete arrays (SPDAs), which consist of separate non-reconfigurabile antenna elements, CAPAs utilize a electromagnetic (EM) surface with a capability of complete analog amplitude and phase control over a continuous aperture. This continuous structure enables precise beamforming capabilities, allowing for enhanced spatial resolution and the ability to approach the upper limits of spatial degrees of freedom within a given spatial resource. The concept of CAPAs can be traced back to current sheets introduced in the 1960s \cite{1138456}. In recent years, several techniques have been proposed to approximate CAPAs, enabling full analog amplitude and phase control by generating high-fidelity RF signals across a near-continuous array of radiating elements, such as metasurface antennas \cite{shlezinger2021dynamic, 10480441, 10908611, castellanos2025embracing, 10981846}, optically driven tightly coupled arrays \cite{8000624}, and interdigital transducer-based grating antennas \cite{yuan2024interdigital}.} Furthermore, the principles underlying CAPA technology are deeply rooted in electromagnetic information theory (EIT), an interdisciplinary framework that integrates EM theory and information theory to analyze physical systems for wireless communication. By leveraging EIT, CAPAs can be optimized to harness the full potential of EM fields, leading to more efficient and higher-capacity communication systems.

\subsection{Prior Works}

Early studies on CAPA technology focused on characterizing the spatial degrees-of-freedom (DoFs) between two CAPAs (sometimes referred to as two volumes) based on EM theory. In particular, the authors of \cite{bucci1989degrees} established the relationship between the DoFs of scattered EM fields and Nyquist sampling principles. A rigorous method for DoF characterization was proposed by the authors of \cite{miller2000communicating}, who derived the proportionality of communication DoFs to the physical volumes of transmit and receive CAPAs through eigenfunction analysis. Subsequent studies refined these insights, analyzing DoFs in various scenarios and providing more analytical expressions \cite{1386525, dardari2020communicating,decarli2021communication, 9798854, 9896943, 9848802, 10262267}. Channel capacity is another important issue for CAPA systems and has attracted significant research attention. Due to the continuous nature of the signal model in CAPA systems, the conventional methods for characterizing the channel capacity in SPDA systems are no longer applicable. Consequently, numerous advanced methods have been developed to accurately quantify channel capacity for point-to-point CAPA-MIMO systems, including the eigenfunction method \cite{jensen2008capacity}, Kolmogorov information theory \cite{8585146}, Fredholm determinant analysis \cite{wan2023mutual}, and physics-informed algorithms \cite{10012689}. There have also been some initial efforts in characterizing the capacity for the CAPA-aided multiple-input single-output (MISO) broadcast channel and single-input multiple-output (SIMO) multiple access channel \cite{zhao2024continuous}.

In recent years, initial research efforts have emerged to design practical beamforming methods to approach channel capacity. The beamforming design for CAPA-MISO and CAPA-SIMO systems has attracted particular attention due to their simpler channel structures. For example, the authors of \cite{zhang2023pattern} proposed a continuous-to-discrete approximation approach to maximize the sum achievable rate. Exploiting the Fourier-based approach, beamforming optimization for CAPA-SIMO multiple access systems was further investigated in \cite{10612761}. As a further advance, the authors of \cite{10910020} and \cite{10938678} proposed to exploit the calculus of variations (CoV) method to address the continuous signal model in CAPA-MISO broadcast systems directly. This method not only clarified the optimal beamformer structures but also significantly reduced computational complexity compared to the Fourier-based method. Similarly, the optimal beamforming design for CAPA-SIMO multiple access systems was derived in \cite{ouyang2024performance}. The application of deep learning for designing CAPA-MISO beamforming was explored in \cite{guo2024deep}, where a single-subspace method was exploited to achieve the continuous-to-discrete transformation. Additionally, for point-to-point CAPA-MIMO systems, the authors of \cite{9906802} introduced a wavenumber-domain beamforming design based on the Fourier-based approach, effectively approaching channel capacity under line-of-sight (LoS) conditions.

\subsection{Motivation and Contributions}

Beamforming design has been shown to be important for maximizing communication performance and approaching channel capacity in CAPA systems. Due to the continuous nature of the CAPA signal model, discretization approaches are commonly employed to approximate the continuous optimization problem as a discrete one, enabling the use of classical optimization tools. The most representative discretization technique is the Fourier-based approach, which leverages the bandlimited property of spatial signals and half-wavelength sampling in the wavenumber domain to achieve highly accurate discretization \cite{9724113, zhang2023pattern, 9906802}. However, the discretization method typically encounters two practical challenges. The first is the inevitable discretization loss compared to the original continuous model. The second is the high dimensionality of the discretized signal. For instance, the number of discretization points required by the Fourier-based approach grows dramatically with increasing CAPA aperture size and signal frequency, leading to substantial computational complexity.
 
Although several techniques, such as the CoV method \cite{10910020, 10938678, ouyang2024performance}, have been proposed to overcome the challenges associated with the Fourier-based method, these solutions are primarily tailored for CAPA-MISO or CAPA-SIMO systems and cannot be directly applied to the considerably more complex CAPA-MIMO channel. Nevertheless, the idea of the CoV method \cite{10910020, 10938678, ouyang2024performance} is particularly appealing, as it directly addresses the continuous signal model without requiring discretization approximation. This method not only enhances communication performance but also significantly reduces computational complexity. These promising observations motivate us to directly tackle the continuous beamforming design for CAPA-MIMO systems without relying on discretization approximations—a problem for which, to the best of our knowledge, no existing method currently exists. The contributions of this paper can be summarized as follows:
\begin{itemize}
    \item We study the beamforming design for a point-to-point uni-polarized CAPA-MIMO system. For the first time, we derive an explicit closed-form expression for the achievable rate as a function of transmit beamformers. 
    \item We propose an iterative weighted minimum mean-squared error (WMMSE) algorithm to directly solve the continuous beamforming optimization problem for maximizing the achievable rate without relying on discretization approximations. Specifically, we characterize the rate-MMSE relationship for CAPA-MIMO and derive closed-form continuous receivers and beamformers at each iteration by employing the CoV. Furthermore, we introduce an equivalent, matrix-based implementation of the WMMSE algorithm using Gauss-Legendre quadrature, which facilitates practical usage.
    \item We present comprehensive numerical results demonstrating the effectiveness of the proposed algorithm. Key findings include: 1) CAPA-MIMO substantially enhances the achievable rate compared to SPAD-MIMO; 2) the proposed WMMSE algorithm improves the achievable rate while significantly reducing computational complexity compared to the conventional Fourier-based method; and 3) the proposed WMMSE algorithm can achieve nearly parallel, non-interference transmission of multiple data streams.
\end{itemize}

\subsection{Organization and Notations}
The remainder of this paper is structured as follows. Section \ref{sec:model} introduces the system model and formulates the beamforming design problem for CAPA-MIMO. Section \ref{sec:algorithm} details the proposed WMMSE algorithm. Section \ref{sec:Fourier} provides a comparison between the proposed methods and the state-of-the-art Fourier-based approach. Numerical results evaluating the performance of different methods under various system configurations are presented in Section \ref{sec:results}. Finally, Section \ref{sec:conclusion} concludes the paper.

\emph{Notations:} Scalars, vectors/matrices, and Euclidean subspaces are denoted by regular, boldface, and calligraphic letters, respectively. The sets of complex, real, and integer numbers are represented by $\mathbb{C}$, $\mathbb{R}$, and $\mathbb{Z}$, respectively. The inverse, transpose, conjugate transpose (Hermitian transpose), and trace operations are represented by $(\cdot)^{-1}$, $(\cdot)^T$, $(\cdot)^H$, and $\mathrm{Tr}(\cdot)$, respectively. The absolute value and Euclidean norm are indicated by $|\cdot|$ and $\|\cdot\|$, respectively. The ceiling function is denoted by $\lceil \cdot \rceil$. The expectation operator is denoted by $\mathbb{E}[\cdot]$. \textcolor{black}{The Lebesgue measure of a Euclidean subspace $\mathcal{S}$ is denoted by $|\mathcal{S}|$.} The real part of a complex number is denoted by $\Re {\cdot}$. An identity matrix of size $N \times N$ is denoted by $\mathbf{I}_N$. Big-O notation is represented by $O(\cdot)$. $\mathrm{rect}(\mathbf{s} \in \mathcal{S})$ denotes the rectangular function, which equals 1 if $\mathbf{s} \in \mathcal{S}$ and 0 otherwise
\section{System Model and Problem Formulation} \label{sec:model}


\subsection{System Model}
\begin{figure}[t!]
    \centering
    \includegraphics[width=0.48\textwidth]{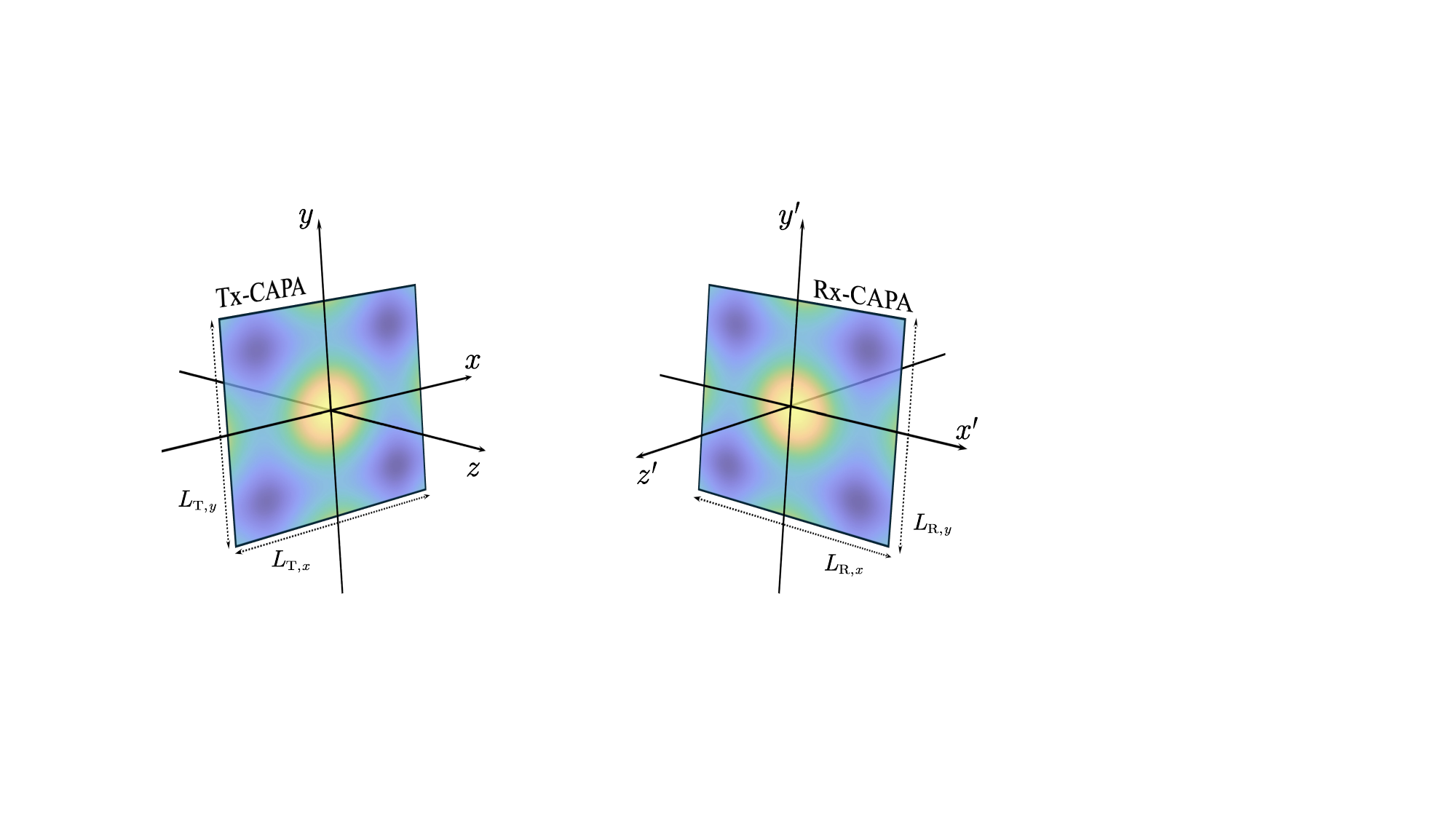}
    \caption{Illustration of the point-to-point CAPA-MIMO system.}
    \label{system_model}
\end{figure}

\textcolor{black}{As shown in Fig. \ref{system_model}, we consider a narrowband single-carrier MIMO system with CAPAs equipped at both transceivers.} The continuous surfaces of the Tx- and Rx-CAPAs are denoted by $\mathcal{S}_{\mathrm{T}}$ and $\mathcal{S}_{\mathrm{R}}$, respectively. Without loss of generality, the square-shape Tx-CAPA is placed on the $x \textendash y$ plane, centered at the origin. Its two sides are aligned parallel to the $x$- and $y$-axes, with lengths along these axes denoted by $L_{\mathrm{T},x}$ and $L_{\mathrm{T},y}$, respectively. Under this circumstance, the coordinate of a point on the Tx-CAPA can be expressed as $\mathbf{s} = [s_x, s_y, 0]^T$. To model the coordinates of a point on the Rx-CAPA, we introduce an auxiliary coordinate system in which the Rx-CAPA is centered at the origin. Its two sides are aligned parallel to the new $x$- and $y$-axes, with lengths along these axes denoted by $L_{\mathrm{R},x}$ and $L_{\mathrm{R},y}$, respectively. Let $\hat{\mathbf{r}} = [ \hat{r}_x, \hat{r}_y, 0]^T$ denote the coordinate of a point on the Rx-CAPA in the auxiliary coordinate system. The coordinates of this point in the original coordinate system can then be expressed as
\begin{equation}
    \mathbf{r} = \mathbf{R}_z(\alpha) \mathbf{R}_y(\beta) \mathbf{R}_x(\phi)  \hat{\mathbf{r}} + \mathbf{r}_o,
\end{equation} 
where $\alpha$, $\beta$, and $\phi$ are angles that the auxiliary coordinate system rotates along the $z$-, $y-$, and $x$-axes of the original coordinate system, respectively, the matrices $\mathbf{R}_z(\alpha)$, $\mathbf{R}_y(\beta)$, and $\mathbf{R}_x(\phi)$ are corresponding rotation matrices, and $\mathbf{r}_o$ is the coordinate of the origin of the auxiliary coordinate system in the original coordinate system. 

The Tx-CAPA contains sinusoidal source currents to radiate information-bearing EM waves. Let $\mathbf{J}(\mathbf{s}) \in \mathbb{C}^{3 \times 1}, \forall \mathbf{s} \in \mathcal{S}_{\mathrm{T}}$, denote the Fourier transform of the source current density at the signal frequency $f$. The electric filed density generated by the source current at the Rx-CAPA can be expressed as 
\begin{equation}
    \mathbf{E}(\mathbf{r}) = \int_{\mathcal{S}_{\mathrm{T}}} \overline{\overline{\mathbf{G}}}(\mathbf{r}, \mathbf{s}) \mathbf{J}(\mathbf{s}) d \mathbf{s} \in \mathbb{C}^{3 \times 1}, \forall \mathbf{r} \in \mathcal{S}_{\mathrm{R}},
\end{equation}     
where $\overline{\overline{\mathbf{G}}}(\mathbf{r}, \mathbf{s}) \in \mathbb{C}^{3 \times 3}$ is the tensor Green's function. Under the LoS condition, the Green's function can be derived from the inhomogeneous Helmholtz wave equation as\footnote{\textcolor{black}{In this paper, we consider a continuous LoS channel within a homogeneous EM medium, as in \cite{zhang2023pattern} and \cite{9906802}, to facilitate an effective and fair comparison with the methods proposed therein. However, it is important to note that the methods we propose in the following section do not rely on any particular structural assumptions about the continuous channel and can therefore be readily exploited to address beamforming problems in more realistic scenarios, incorporating factors such as inhomogeneous EM media, antenna efficiency, impedance matching, and multipath fading.}} \cite{dardari2020communicating} 
\begin{equation}
    \overline{\overline{\mathbf{G}}}(\mathbf{r}, \mathbf{s}) = \frac{-j \eta e^{-j \frac{2\pi}{\lambda}\|\mathbf{r} - \mathbf{s}\|}}{2 \lambda \|\mathbf{r}-\mathbf{s}\|} \left( \mathbf{I}_3 - \frac{(\mathbf{r}-\mathbf{s})(\mathbf{r}-\mathbf{s})^T}{\|\mathbf{r} - \mathbf{s}\|^2} \right),
\end{equation}
where $\eta$ denotes the free-space impedance, $\lambda = c/f$ denotes the signal wavelength with $c$ representing the speed of light. \textcolor{black}{It is beneficial to exploit multiple polarization states, i.e., different dimensions of $\mathbf{J}(\mathbf{s})$ and $\mathbf{E}(\mathbf{r})$ for enhancing communication performance, but it will also lead to higher design and hardware complexity. Therefore, this study focus on uni-polarized CAPAs, with the polarization direction of the Tx-CAPA aligned along the $y$-axis.} Under this configuration, the source current density is expressed as 
\begin{equation}
    \mathbf{J}(\mathbf{s}) = x(\mathbf{s}) \mathbf{u}_{\mathrm{T}},
\end{equation} 
where $\mathbf{u}_{\mathrm{T}} = [0,1,0]^T$ represents the polarization direction vector and $x(\mathbf{s}) \in \mathbb{C}$ is the corresponding component of the source current. Assuming $N$ independent data streams to transmit, $x(\mathbf{s})$ can be expressed as 
\begin{equation} \label{expression_transmit_signal}
    x(\mathbf{s}) = \sum_{n=1}^N w_n(\mathbf{s}) c_n = \mathbf{w}(\mathbf{s}) \mathbf{c},
\end{equation}
where $\mathbf{w}(\mathbf{s}) = [w_1(\mathbf{s}),\dots,w_N(\mathbf{s})] \in \mathbb{C}^{1 \times N}$ and $\mathbf{c} = [c_1,\dots,c_N]^T \in \mathbb{C}^{N \times 1}$, with $w_n (\mathbf{s}) \in \mathbb{C}$ denoting the continuous beamformer for the $n$-th data stream $c_n \in \mathbb{C}$. The data streams are assumed to be independent and have unit average power, satisfying $\mathbb{E}[ \mathbf{c} \mathbf{c}^H ] = \mathbf{I}_N$.

\textcolor{black}{Similarly, the polarization direction of the Rx-CAPA aligns with the $y$-axis of its local auxiliary coordinate system, leading to a polarization vector $\mathbf{u}_{\mathrm{R}} = \mathbf{R}_z(\alpha) \mathbf{R}_y(\beta) \mathbf{R}_x(\phi) \hat{\mathbf{u}}_{\mathrm{R}}$, where $\hat{\mathbf{u}}_{\mathrm{R}} = [0,1,0]^T$.} Therefore, the effective noisy electric field density observed by the Rx-CAPA is given by \cite{9906802, zhang2023pattern}
\begin{equation}
    y(\mathbf{r}) = \mathbf{u}_{\mathrm{R}}^T \mathbf{E}(\mathbf{r}) + z(\mathbf{r}) = \int_{\mathcal{S}_{\mathrm{T}}} h(\mathbf{r}, \mathbf{s}) x(\mathbf{s}) d \mathbf{s}+ z(\mathbf{r}),
\end{equation}    
where $h(\mathbf{r}, \mathbf{s}) = \mathbf{u}_{\mathrm{R}}^T \overline{\overline{\mathbf{G}}}(\mathbf{r}, \mathbf{s}) \mathbf{u}_{\mathrm{T}} \in \mathbb{C}$ denotes the channel response and $z(\mathbf{r}) \sim \mathcal{CN}(0, \sigma^2)$ denotes additive white Gaussian noise\footnote{\textcolor{black}{For simplicity, we assume white Gaussian noise here, which can be attributed to measurement noise \cite{4685903, wan2023mutual}. However, when accounting for radiation interference, the noise may become spatially correlated \cite{9906802, wan2023mutual}. }}.


\subsection{Problem Formulation}

In this paper, we aim to optimize the transmit beamformer to maximize the achievable rate. Although the achievable rate of CAPA-MIMO has been extensively studied in the literature, most existing works rely on approximations and bounding techniques and, therefore, do not provide an explicit and accurate expression of the achievable rate as a function of the transmit beamformer. To bridge this gap, we formally characterize the achievable rate in the following theorem.

\begin{theorem} \label{theorem_SU_capacity}
    \normalfont
    \emph{(Achievable rate of CAPA-MIMO)} The achievable rate that the data streams can reliably carry through channel $h(\mathbf{r}, \mathbf{s})$ with the beamformer $\mathbf{w}(\mathbf{s})$ is given by 
    \begin{align} \label{SU_capacity}
        R = \log \det \left( \mathbf{I}_N + \frac{1}{\sigma^2} \mathbf{Q} \right),
    \end{align}
    where 
    \begin{align}
        \label{expression_of_e}
        \mathbf{e}(\mathbf{r}) &= \int_{\mathcal{S}_{\mathrm{T}}} h(\mathbf{r}, \mathbf{s}) \mathbf{w}(\mathbf{s}) d \mathbf{s}, \\
        \label{expression_of_Q}
        \mathbf{Q} &= \int_{\mathcal{S}_{\mathrm{R}}} \mathbf{e}^H(\mathbf{r}) \mathbf{e}(\mathbf{r}) d \mathbf{r}.
    \end{align}
\end{theorem}

\begin{IEEEproof}
    Please refer to Appendix \ref{theorem_SU_capacity_proof}
\end{IEEEproof}

\textcolor{black}{
\begin{remark}
    \normalfont
    MIMO communication requires beamformer design at both the transmitter and receiver. Although the achievable rate expression in \eqref{SU_capacity} depends explicitly only on the transmit beamformer, the receive beamformer is implicitly accounted for. As shown in the proof in Appendix \ref{theorem_SU_capacity_proof}, the optimal receive beamformer corresponding to any given transmit beamformer is the MMSE receiver. Therefore, in the following, we focus on the design of the transmit beamformer.
\end{remark}
}

\textcolor{black}{
\begin{remark}
    \normalfont
    The achievable rate expression in \textbf{Theorem \ref{theorem_SU_capacity}} aligns with the classical result for conventional SPDAs. Let $\mathbf{H}_d \in \mathbb{C}^{M_{\mathrm{R}} \times M_{\mathrm{T}}}$ denote the point-to-point MIMO channel with SPDAs, with $M_{\mathrm{R}}$ and $M_{\mathrm{T}}$ being the number of discrete antennas at the Tx and Rx, respectively. Then, under the Guassian signalling assumption, the achievable rate can be expressed as 
    \begin{align} \label{SU_capacity_SPDA_1}
        R_d = &\log \det \left( \mathbf{I}_{M_{\mathrm{R}}} + \frac{1}{\sigma^2} \mathbf{H}_d \mathbf{W}_d \mathbf{W}_d^H \mathbf{H}_d^H \right) \nonumber \\
        = &\log \det \left( \mathbf{I}_N + \frac{1}{\sigma^2} \mathbf{W}_d^H \mathbf{H}_d^H \mathbf{H}_d \mathbf{W}_d \right),
    \end{align}
    where $\mathbf{W}_d \in \mathbb{C}^{M_{\mathrm{T}} \times N}$ denotes the discrete transmit beamformer. Comparing \eqref{SU_capacity_SPDA_1} with \eqref{SU_capacity}, we observe that the matrix multiplications along the $M_{\mathrm{T}}$ and $M_{\mathrm{R}}$ dimensions in the SPDA model correspond to integrals over $\mathcal{S}_{\mathrm{T}}$ and $\mathcal{S}_{\mathrm{R}}$ in the CAPA model. In particular, the expression $\mathbf{e}(\mathbf{r})$ for CAPAs corresponds to the expression $\mathbf{H}_d \mathbf{W}_d$ for SPDAs.    
\end{remark}
}

Based on \textbf{Theorem \ref{theorem_SU_capacity}}, the achievable rate maximization problem can be formulated as
\begin{subequations} \label{problem_1}
    \begin{align}
        \max_{\mathbf{w}(\mathbf{s})} \quad &\log \det \left( \mathbf{I}_N + \frac{1}{\sigma^2} \mathbf{Q} \right) \\
        \label{power_constraint}
        \mathrm{s.t.} \quad & \int_{\mathcal{S}_{\mathrm{T}}} \|\mathbf{w}(\mathbf{s})\|^2 d\mathbf{s} \le P_{\mathrm{T}},
    \end{align}
\end{subequations} 
where \eqref{power_constraint} \textcolor{black}{limits the transmit power to within the budget $P_{\mathrm{T}}$~\cite{9906802}. }

\subsection{The “Elusive” Optimal Solution}
Theoretically, the optimal continuous beamformer that maximizes the achievable rate is “easy” to derive. To elaborate, we rewrite the achievable rate as
\begin{align} \label{K_achievable_rate}
    R = \log \det \left( \mathbf{I}_N + \frac{1}{\sigma^2} \int_{\mathcal{S}_{\mathrm{T}}} \int_{\mathcal{S}_{\mathrm{T}}} \mathbf{w}^H(\mathbf{s}) K(\mathbf{s}, \mathbf{z}) \mathbf{w}(\mathbf{z}) d \mathbf{z} d \mathbf{s}  \right),
\end{align}
where $K(\mathbf{s}, \mathbf{z})$ is a Hermitian kernel operator measuring the coupling between the two points $\mathbf{s}$ and $\mathbf{z}$, given by 
\begin{equation}
    K(\mathbf{s}, \mathbf{z}) = \int_{\mathcal{S}_{\mathrm{R}}} h^H(\mathbf{r}, \mathbf{s}) h(\mathbf{r}, \mathbf{z}) d \mathbf{r}.
\end{equation} 
\textcolor{black}{Similar to the eigendecomposition of a matrix, the kernel $K(\mathbf{s}, \mathbf{z})$ can be decomposed into a complete set of basis functions \cite{miller2000communicating, 9906802}:
\begin{equation} \label{eigendecomposition_K}
    K(\mathbf{s}, \mathbf{z}) = \sum_{n=1}^{\infty} \xi_n \phi_n(\mathbf{s}) \phi_n^H(\mathbf{z}),
\end{equation}
where $\xi_n$ is the non-negative eigenvalue and the $\phi_n(\mathbf{s})$ is the orthogonal eigenfunction. It follows that 
\begin{equation}
    \xi_n \phi_n(\mathbf{s}) = \int_{\mathcal{S}_{\mathrm{T}}} K(\mathbf{s}, \mathbf{z}) \phi_n(\mathbf{z}) d \mathbf{z}.  
\end{equation}
Therefore, assuming $\xi_1 \ge \xi_2 \ge \dots \ge \xi_n \ge \dots$, the optimal beamformer can be obtained as \cite{miller2000communicating, 9906802}
\begin{equation}
    w_n(\mathbf{s}) = \sqrt{P_n} \phi_n(\mathbf{s}), \quad n=1,\dots,N,
\end{equation}
where $P_n$ is a power allocation coefficient that can be obtained using the water-filling strategy, satisfying $\sum_{n=1}^N P_n = P_{\mathrm{T}}$. }

Nevertheless, although deriving the optimal beamformer is straightforward, its practical implementation remains prohibitive due to the extremely high computational complexity associated with numerically solving the eigendecomposition problem \eqref{eigendecomposition_K}. A common state-of-the-art approach is to discretize the continuous kernel using Fourier basis functions \cite{9906802}. However, this approach inevitably incurs discretization losses and leads to high-dimensional discretized kernels. To overcome these limitations, in the following section, we propose a WMMSE algorithm that directly solves the rate maximization problem in \eqref{problem_1}, thus effectively eliminating discretization losses and significantly reducing computational complexity.

\section{WMMSE Algorithm for CAPA-MIMO} \label{sec:algorithm}

This section describes the proposed WMMSE algorithm for solving problem \eqref{problem_1}. While the WMMSE algorithm is typically unnecessary for conventional point-to-point SPDA-MIMO, where the optimal beamformer can be directly obtained via singular-value decomposition (SVD), it becomes particularly useful for point-to-point CAPA-MIMO. Due to the complexity involved in solving the eigendecomposition problem for CAPAs, the WMMSE algorithm provides an effective alternative by circumventing the need for eigendecomposition. In the following, we elaborate on adapting the WMMSE algorithm specifically for CAPA-MIMO systems.


\subsection{WMMSE Algorithm} \label{sec:WMMSE}
To streamline the derivation of the WMMSE algorithm, we first transform problem \eqref{problem_1} into an equivalent unconstrained problem using the following lemma.

\begin{lemma} \label{equal_power_lemma}
    \normalfont
    \emph{(Equivalent unconstrained problem)}
    Let $\mathbf{w}^{\dagger}(\mathbf{s})$ denote the optimal solution to the following unconstrained optimization problem:
    \begin{equation} \label{unconstrained_problem}
        \max_{\mathbf{w}(\mathbf{s})} \quad \widetilde{R} = \log \det \left( \mathbf{I}_N + \frac{1}{\widetilde{\sigma}^2} \mathbf{Q} \right),
    \end{equation}
    where 
    \begin{equation}
        \label{expression_sigma}
        \widetilde{\sigma}^2 = \frac{\sigma^2}{P_{\mathrm{T}}} \int_{\mathcal{S}_{\mathrm{T}}} \|\mathbf{w}(\mathbf{s})\|^2 d\mathbf{s}.
    \end{equation}
    The optimal solution to problem \eqref{problem_1} is given by
    \begin{equation} \label{scale_w}
        \mathbf{w}^\star(\mathbf{s}) = \sqrt{\frac{P_{\mathrm{T}}}{\int_{\mathcal{S}_{\mathrm{T}}} \|\mathbf{w}^{\dagger}(\mathbf{s})\|^2 d\mathbf{s}}} \mathbf{w}^{\dagger}(\mathbf{s}).
    \end{equation}
\end{lemma}

\begin{IEEEproof}
    It can be readily verified that the optimal solution to problem \eqref{problem_1} must satisfy the power constraint with equality, a condition clearly met by the solution provided in \eqref{scale_w}. Furthermore, by substituting \eqref{scale_w} into \eqref{problem_1}, one can verify that the resulting objective value is always equal to the objective value of \eqref{unconstrained_problem}. This completes the proof.
\end{IEEEproof}

\textcolor{black}{Based on \textbf{Lemma \ref{equal_power_lemma}}, we focus on solving the unconstrained problem \eqref{unconstrained_problem} in the following. The corresponding signal model for the new achievable rate $\widetilde{R}$  in problem \eqref{unconstrained_problem} is given by
\begin{equation}
    \widetilde{y}(\mathbf{r}) = \int_{\mathcal{S}_{\mathrm{T}}} h(\mathbf{r}, \mathbf{s}) x(\mathbf{s}) d \mathbf{s} + \widetilde{z}(\mathbf{r}),
\end{equation}
where $z(\mathbf{r}) \sim \mathcal{CN}(0, \widetilde{\sigma}^2)$ is the equivalent noise. Note that in this new model, the original noise power is replaced by the scaled noise power given in \eqref{expression_sigma}, allowing the power constraint to be safely omitted.}

The key idea of the WMMSE algorithm is to transform the rate-maximization problem into an equivalent weighted-MSE minimization problem \cite{4712693,shi2011iteratively}. Let $\mathbf{v}(\mathbf{r}) \in \mathbb{C}^{1 \times N}$ denote the receiver to estimate the data stream $\mathbf{c}$ from $\widetilde{y}(\mathbf{r})$. The estimated signal is given by 
\begin{align}
    \hat{\mathbf{c}} = &\int_{\mathcal{S}_{\mathrm{R}}} \mathbf{v}^H (\mathbf{r}) \widetilde{y}(\mathbf{r}) d \mathbf{r} 
    = \int_{\mathcal{S}_{\mathrm{R}}} \mathbf{v}^H(\mathbf{r}) \mathbf{e}(\mathbf{r}) \mathbf{c} d \mathbf{r} + \widetilde{\mathbf{z}},
\end{align}   
where 
\begin{align}
    \widetilde{\mathbf{z}} = \int_{\mathcal{S}_{\mathrm{R}}} \mathbf{v}^H(\mathbf{r}) \widetilde{z} (\mathbf{r}) d \mathbf{r} \sim \mathcal{CN} \left( \mathbf{0}, \widetilde{\sigma}^2 \int_{\mathcal{S}_{\mathrm{R}}} \mathbf{v}^H(\mathbf{r}) \mathbf{v}(\mathbf{r})  d \mathbf{r}  \right).
\end{align}
The above distribution of $\widetilde{\mathbf{z}}$ can be obtained following  \cite[Lemma 1]{ouyang2024performance}. The MSE matrix related to the receiver $\mathbf{v}(\mathbf{r})$ is given by 
\begin{align} \label{MSE_matrix}
    \mathbf{E} = & \mathbb{E} \left[ (\hat{\mathbf{c}} - \mathbf{c})(\hat{\mathbf{c}} - \mathbf{c})^H  \right] \nonumber \\
    = & \left( \mathbf{I}_N - \int_{\mathcal{S}_{\mathrm{R}}} \mathbf{v}^H(\mathbf{r}) \mathbf{e}(\mathbf{r}) d \mathbf{r} \right) \left( \mathbf{I}_N - \int_{\mathcal{S}_{\mathrm{R}}} \mathbf{v}^H(\mathbf{r}) \mathbf{e}(\mathbf{r}) d \mathbf{r} \right)^H \nonumber \\
    &\hspace{3.5cm} + \widetilde{\sigma}^2 \int_{\mathcal{S}_{\mathrm{R}}} \mathbf{v}^H(\mathbf{r}) \mathbf{v}(\mathbf{r})  d \mathbf{r}
\end{align}
The optimal MMSE receiver is given by\footnote{The MMSE receiver in \eqref{MMSE_receiver_no_SIC} is derived under the assumption that successive interference cancellation (SIC) is not employed, and thus is different from the MMSE receiver presented in Appendix~\ref{theorem_SU_capacity_proof}, which is derived under the condition that SIC is applied. } \cite[Theorem 4]{ouyang2024performance}
\begin{align} \label{MMSE_receiver_no_SIC}
    \mathbf{v}_{\mathrm{MMSE}}(\mathbf{r}) = &\argmin_{\mathbf{v}(\mathbf{r})} \mathrm{Tr} \left( \mathbf{E} \right)
    = \argmin_{\mathbf{v}(\mathbf{r})} \mathbb{E} \left[ \left\| \hat{\mathbf{c}} - \mathbf{c}  \right\|^2 \right] \nonumber \\
    = & \frac{1}{\widetilde{\sigma}^2} \mathbf{e}(\mathbf{r}) \left( \mathbf{I}_N + \frac{1}{\widetilde{\sigma}^2}  \mathbf{Q} \right)^{-1}.
\end{align}
To obtain the MSE matrix achieved by the MMSE receiver, we first derive the following intermediate expressions:
\begin{align}
    \label{MSE_matrix_1}
    &\mathbf{I}_N - \int_{\mathcal{S}_{\mathrm{R}}} \mathbf{v}_{\mathrm{MMSE}}^H(\mathbf{r}) \mathbf{e}(\mathbf{r}) d \mathbf{r}  \nonumber \\
    & \quad = \mathbf{I}_N - \frac{1}{\widetilde{\sigma}^2} \left(\mathbf{I}_N + \frac{1}{\widetilde{\sigma}^2} \mathbf{Q}\right)^{-1} \mathbf{Q} = \left(\mathbf{I}_N + \frac{1}{\widetilde{\sigma}^2} \mathbf{Q}\right)^{-1}, \\
    \label{MSE_matrix_2}
    & \int_{\mathcal{S}_{\mathrm{R}}} \mathbf{v}_{\mathrm{MMSE}}^H(\mathbf{r}) \mathbf{v}_{\mathrm{MMSE}}(\mathbf{r})  d \mathbf{r} \nonumber \\
    & \quad = \frac{1}{\widetilde{\sigma}^4} \left(\mathbf{I}_N + \frac{1}{\widetilde{\sigma}^2} \mathbf{Q}\right)^{-1} \mathbf{Q} \left(\mathbf{I}_N + \frac{1}{\widetilde{\sigma}^2} \mathbf{Q}\right)^{-1}.
\end{align}
By substituting \eqref{MSE_matrix_1} and \eqref{MSE_matrix_2} into \eqref{MSE_matrix}, the MSE matrix achieved by the MMSE receiver can be obtained as
\begin{align} \label{MMSE_MSE_matrix}
    \mathbf{E}_{\mathrm{MMSE}}
    = &\left(\mathbf{I}_N + \frac{1}{\widetilde{\sigma}^2} \mathbf{Q}\right)^{-1},
\end{align}
\textcolor{black}{By comparing \eqref{unconstrained_problem} and \eqref{MMSE_MSE_matrix}, it is easy to obtain a rate-MMSE relationship of $\widetilde{R} = \log \det \left( \mathbf{E}_{\mathrm{MMSE}}^{-1} \right)$.} Based on this relationship, the rate maximization problem can be equivalently reformulated as an MSE minimization problem, as stated in the following theorem.

\begin{theorem} \label{theorem_2}
    \normalfont 
    \emph{(Equivalent MSE minimization problem)}
    Define a weight matrix $\mathbf{U} \succeq 0$. The optimal solution to problem \eqref{unconstrained_problem} is identical to the following problem:
    \begin{equation} \label{MSE_minimization_problem}
        \min_{\mathbf{w}(\mathbf{s}), \mathbf{v}(\mathbf{r}), \mathbf{U}} \quad \mathrm{Tr} \left( \mathbf{U} \mathbf{E} \right) - \log \det (\mathbf{U}).
    \end{equation} 
\end{theorem}

\begin{IEEEproof}
    Please refer to Appendix \ref{theorem_2_proof}.
\end{IEEEproof}

Based on the equivalence established in \textbf{Theorem \ref{theorem_2}}, the solution of $\mathbf{w}(\mathbf{s})$ to problem \eqref{unconstrained_problem} can be obtained by solving problem \eqref{MSE_minimization_problem} in an alternating manner. Specifically, according to the proof in Appendix \ref{theorem_2_proof}, for any given $\mathbf{w}(\mathbf{s})$, the optimal solutions of $\mathbf{v}(\mathbf{r})$ and $\mathbf{U}$ are given by $\mathbf{v}_{\mathrm{MMSE}}(\mathbf{r})$ and $\mathbf{E}_{\mathrm{MMSE}}^{-1}$, respectively. Furthermore, for any given $\mathbf{v}(\mathbf{r})$ and $\mathbf{U}$, the optimal solution of $\mathbf{w}(\mathbf{s})$ is given in the following proposition. 

\begin{proposition}
    \normalfont \label{optimal_w_s}
    For any given $\mathbf{v}(\mathbf{r})$ and $\mathbf{U}$, the optimal solution of $\mathbf{w}(\mathbf{s})$ to the MSE minimization problem \eqref{MSE_minimization_problem} is given by  
    \begin{equation} \label{optimal_w_function}
        \mathbf{w}(\mathbf{s}) = \mathbf{g}(\mathbf{s}) \mathbf{U}  \left(\frac{1}{\varepsilon} \mathbf{I}_N + \mathbf{G} \mathbf{U} \right)^{-1},
    \end{equation}
    where 
    \begin{align}
        \label{expression_g}
        & \mathbf{g}(\mathbf{s}) = \int_{\mathcal{S}_{\mathrm{R}}} h^H(\mathbf{r}, \mathbf{s}) \mathbf{v}(\mathbf{r}) d \mathbf{r}, \quad  \mathbf{G} = \int_{\mathcal{S}_{\mathrm{T}}} \mathbf{g}^H(\mathbf{s}) \mathbf{g}(\mathbf{s}) d \mathbf{s}, \\
        \label{expression_V}
        & \mathbf{V} = \int_{\mathcal{S}_{\mathrm{R}}} \mathbf{v}^H(\mathbf{r}) \mathbf{v}(\mathbf{r})  d \mathbf{r}, \quad \varepsilon = \frac{P_{\mathrm{T}}}{\sigma^2 \mathrm{Tr}(\mathbf{U} \mathbf{V})}.
    \end{align}


\end{proposition}

\begin{IEEEproof}
    Please refer to Appendix \ref{optimal_w_s_proof}.
\end{IEEEproof}

Based on the above results, the WMMSE algorithm for solving problem \eqref{problem_1} is summarized in \textbf{Algorithm \ref{alg:SU_WMMSE}}.

\begin{algorithm}[tb]
    \caption{WMMSE Algorithm for CAPA-MIMO Achievable Rate Maximization}
    \label{alg:SU_WMMSE}
    \begin{algorithmic}[1]
        \STATE{initialize $t=0$ and $\mathbf{w}_0(\mathbf{s})$}
        \REPEAT
            \STATE{update $\mathbf{v}_{t+1}(\mathbf{r})$ as $\mathbf{v}_{\mathrm{MMSE}}(\mathbf{r})$ for given $\mathbf{w}_t(\mathbf{s})$}
            \STATE{update $\mathbf{U}_{t+1}$ as $\mathbf{E}_{\mathrm{MMSE}}^{-1}$ for given $\mathbf{w}_t(\mathbf{s})$}
            \STATE{update $\mathbf{w}_{t+1}(\mathbf{s})$ by \eqref{optimal_w_function} for given $\mathbf{v}_{t+1}(\mathbf{r})$ and $\mathbf{U}_{t+1}$}
        \UNTIL{the fractional increase of the objective value of problem \eqref{problem_1} falls below a predefined threshold}
        \STATE{scale the converged $\mathbf{w}(\mathbf{s})$ according to \eqref{scale_w}}
    \end{algorithmic}
\end{algorithm}

\subsection{Low-Complexity Implementation}

\textcolor{black}{In the WMMSE algorithm described in \textbf{Algorithm \ref{alg:SU_WMMSE}}, although each optimization variable update has a closed-form solution, its practical implementation remains challenging due to the need for computing integrals and handling continuous functions in each iteration.} To overcome this challenge, we propose a practical matrix-based implementation of the WMMSE algorithm using Gauss-Legendre quadrature, which is a widely-used effective numerical integration method and takes the form \cite{olver2010nist}:
\begin{equation} \label{GL_quadrature}
    \int_{a}^{b} f(x) d x \approx \frac{b-a}{2} \sum_{m=1}^M \omega_m f \left( \frac{b-a}{2} \theta_m + \frac{a+b}{2} \right),
\end{equation} 
where $M$ is the number of sample points, $\{\omega_m\}_{m=1}^M$ are the quadrature weights, and $\{\theta_m\}_{m=1}^M$ are the roots of the $M$-th Legendre polynomial. The Gauss-Legendre quadrature converges geometrically as $M$ increases. Therefore, a small value of $M$, typically between 10 and 20, is usually sufficient to obtain the near-exact integral results. Therefore, in the following, we assume that the value of $M$ is selected such that \eqref{GL_quadrature} holds almost with equality.

In the CAPA-MIMO system, there are two kinds of integrals that need to be calculated, i.e., the integrals over $\mathcal{S}_{\mathrm{T}}$ and $\mathcal{S}_{\mathrm{R}}$, respectively. By using the Gauss-Legendre quadrature, the integral over $\mathcal{S}_{\mathrm{T}}$ can be reformulated as 
\begin{align} \label{GL_Tx}
    \int_{\mathcal{S}_{\mathrm{T}}} f(\mathbf{s}) d \mathbf{s} = &\int_{-\frac{L_{\mathrm{T},x}}{2}}^{\frac{L_{\mathrm{T}, x}}{2}} \int_{-\frac{L_{\mathrm{T}, y}}{2}}^{\frac{L_{\mathrm{T}, y}}{2}} f(\mathbf{s})  d s_x d s_y \nonumber \\
    = & \sum_{n=1}^{M} \sum_{m=1}^{M} \frac{\omega_n \omega_m A_{\mathrm{T}}}{4}  f(\mathbf{s}_{n,m}), 
\end{align}  
where 
\begin{align} \label{expression_s_ij}
    A_{\mathrm{T}} &= L_{\mathrm{T}, x} L_{\mathrm{T}, y}, \quad \mathbf{s}_{n,m} = \left[ \frac{\theta_i L_{\mathrm{T}, x}}{2}, \frac{\theta_j L_{\mathrm{T}, y}}{2}, 0 \right]^T.
\end{align}
The integral over $\mathcal{S}_{\mathrm{R}}$ can be formulated as 
\begin{align} \label{GL_Rx}
    \int_{\mathcal{S}_{\mathrm{R}}} f(\mathbf{r}) d \mathbf{r} = &\int_{-\frac{L_{\mathrm{R},x}}{2}}^{\frac{L_{\mathrm{R}, x}}{2}} \int_{-\frac{L_{\mathrm{R}, y}}{2}}^{\frac{L_{\mathrm{R}, y}}{2}} f(\mathbf{r}) d \hat{r}_x d \hat{r}_y \nonumber \\
    = &\sum_{n=1}^{M} \sum_{m=1}^{M} \frac{\omega_n \omega_m A_{\mathrm{R}}}{4} f(\mathbf{r}_{n,m}),
\end{align}
where 
\begin{align}
    & A_{\mathrm{R}} = L_{\mathrm{R}, x} L_{\mathrm{R}, y}, \quad \hat{\mathbf{r}}_{n,m} = \left[ \frac{\theta_i L_{\mathrm{R},x}}{2}, \frac{\theta_j L_{\mathrm{R},y}}{2}, 0 \right]^T \\
    & \mathbf{r}_{n,m} = \mathbf{R}(\alpha) \mathbf{R}(\beta) \mathbf{R}(\phi)  \hat{\mathbf{r}}_{n,m} + \mathbf{r}_o.
\end{align}
\textcolor{black}{It can be observed from \eqref{GL_Tx} and \eqref{GL_Rx} that the Gauss-Legendre quadrature effectively transform the continuous integral calculation into the discrete calculation.} These transformations facilitate the matrix-based implementation of the WMMSE algorithm in \textbf{Algorithm \ref{alg:SU_WMMSE}}. In particular, the values of $\mathbf{v}_{t+1}(\mathbf{r})$, $\mathbf{U}_{t+1}$, and $\mathbf{w}_{t+1}(\mathbf{s})$ in the $(t+1)$-th iteration of the WMMSE algorithm need to be calculated by giving the value of $\mathbf{w}_t (\mathbf{s})$ obtained in the $t$-th iteration. Before proceeding to derive these quantities, we first define $M_s = M^2$ and the following notations:
\begin{subequations}
    \begin{align}
        &\mathbf{W}_t = \big[ \mathbf{w}_t^T(\mathbf{s}_{1,1}),\dots,\mathbf{w}_t^T(\mathbf{s}_{M,M})\big]^T \in \mathbb{C}^{M_s \times N}, \\
        &\mathbf{h}(\mathbf{r}) = \big[ h(\mathbf{r}, \mathbf{s}_{1,1}),\dots, h(\mathbf{r}, \mathbf{s}_{M,M})\big] \in \mathbb{C}^{1 \times M_s}, \\
        & \widetilde{\mathbf{h}}(\mathbf{s}) = \big[ h^H(\mathbf{r}_{1,1}, \mathbf{s}),\dots,h^H(\mathbf{r}_{M,M}, \mathbf{s})\big] \in \mathbb{C}^{1 \times M_s}, \\
        \label{GL_channel}
        & \mathbf{H} = \big[\mathbf{h}^T(\mathbf{r}_{1,1}),\dots,\mathbf{h}^T(\mathbf{r}_{M,M})\big]^T \in \mathbb{C}^{M_s \times M_s}, \\
        &\mathbf{\Phi}_{\mathrm{T}} = \frac{1}{4} A_{\mathrm{T}} \mathrm{diag}\left( \omega_1 \omega_1,\dots,\omega_{M} \omega_{M} \right) \in \mathbb{C}^{M_s \times M_s}, \\
        & \mathbf{\Phi}_{\mathrm{R}} = \frac{1}{4} A_{\mathrm{R}} \mathrm{diag}\left( \omega_1 \omega_1,\dots,\omega_{M} \omega_{M} \right) \in \mathbb{C}^{M_s \times M_s}.
    \end{align}
\end{subequations}

\subsubsection{Calculate  $\mathbf{v}_{t+1}(\mathbf{r})$ and $\mathbf{U}_{t+1}$} To calculate these two variables, $\mathbf{e}_{t+1}(\mathbf{r})$, $\widetilde{\sigma}_{t+1}^2$, and $\mathbf{Q}_{t+1}$ need to be calculated according to \eqref{MMSE_receiver_no_SIC} and \eqref{MMSE_MSE_matrix}. In particular, based on the expression \eqref{expression_of_e} and the transformation \eqref{GL_Tx}, $\mathbf{e}_{t+1}(\mathbf{r})$ can be calculated by  
\begin{align}
    \mathbf{e}_{t+1}(\mathbf{r}) 
    = & \sum_{n=1}^{M} \sum_{m=1}^{M} \frac{\omega_n \omega_m A_{\mathrm{T}}}{4} h(\mathbf{r}, \mathbf{s}_{n,m}) \mathbf{w}_t(\mathbf{s}_{n,m}) \nonumber \\
    = &\mathbf{h}(\mathbf{r}) \mathbf{\Phi}_{\mathrm{T}} \mathbf{W}_t.
\end{align}
Based on the expression \eqref{expression_sigma} and the transformation \eqref{GL_Tx}, $\widetilde{\sigma}_{t+1}^2$ can be calculated by  
\begin{align}
    \widetilde{\sigma}_{t+1}^2 
    = &\frac{\sigma^2}{P_{\mathrm{T}}} \sum_{n=1}^{M} \sum_{m=1}^{M} \frac{\omega_n \omega_m A_{\mathrm{T}}}{4} \|\mathbf{w}_t(\mathbf{s}_{n,m})\|^2 \nonumber \\ 
    = &\frac{\sigma^2}{P_{\mathrm{T}}} \mathrm{Tr}\left( \mathbf{W}_t^H \mathbf{\Phi}_{\mathrm{T}} \mathbf{W}_t \right).
\end{align}
Based on the expression \eqref{expression_of_Q} and the transformation \eqref{GL_Rx}, $\mathbf{Q}_{t+1}$ can be calculated by 
\begin{align} \label{expression_Q_t}
    \mathbf{Q}_{t+1} 
    = & \sum_{n=1}^{M} \sum_{m=1}^{M} \frac{\omega_n \omega_m A_{\mathrm{R}}}{4} \mathbf{e}_{t+1}^H(\mathbf{r}_{n,m}) \mathbf{e}_{t+1}(\mathbf{r}_{n,m}) \nonumber \\
    = & \mathbf{W}_t^H \mathbf{\Phi}_{\mathrm{T}} \mathbf{H}^H \mathbf{\Phi}_{\mathrm{R}} \mathbf{H} \mathbf{\Phi}_{\mathrm{T}} \mathbf{W}_t.
\end{align}
Therefore, based on \eqref{MMSE_receiver_no_SIC}, $\mathbf{v}_{t+1}(\mathbf{r})$ can be calculated as  
\begin{align} \label{expression_v_t}
    \mathbf{v}_{t+1}(\mathbf{r}) = & \frac{1}{\widetilde{\sigma}_{t+1}^2} \mathbf{e}_{t+1}(\mathbf{r}) \left( \mathbf{I}_N + \frac{1}{\widetilde{\sigma}_{t+1}^2}  \mathbf{Q}_{t+1} \right)^{-1} \nonumber \\
    = & \mathbf{h}(\mathbf{r}) \mathbf{\Phi}_{\mathrm{T}} \mathbf{W}_t \mathbf{\Theta}_{t+1}^{-1},
\end{align}
where $\mathbf{\Theta}_{t+1} =  \widetilde{\sigma}_{t+1}^2 \mathbf{I}_N + \mathbf{Q}_{t+1}$.
Furthermore, based on \eqref{MMSE_MSE_matrix}, $\mathbf{U}_{t+1}$ can be calculated as 
\begin{equation} \label{expression_U_t}
   \mathbf{U}_{t+1} = \mathbf{I}_N + \frac{1}{\widetilde{\sigma}_{t+1}^2} \mathbf{Q}_{t+1}.
\end{equation} 

\subsubsection{Calculate $\mathbf{w}_{t+1}(\mathbf{s})$} According to \eqref{optimal_w_function}, $\mathbf{g}_{t+1}(\mathbf{s})$, $\mathbf{G}_{t+1}$, and $\mathbf{V}_{t+1}$ need to be calculated first. In particular, based on the expression in \eqref{expression_g} and the transformation \eqref{GL_Rx}, $\mathbf{g}_{t+1}(\mathbf{s})$ can be calculated by 
\begin{align} \label{expression_g_discrete}
    \mathbf{g}_{t+1}(\mathbf{s}) 
    = &\sum_{n=1}^M \sum_{m=1}^M \frac{\omega_n \omega_m A_{\mathrm{R}}}{4} h^H(\mathbf{r}_{n,m}, \mathbf{s}) \mathbf{v}_{t+1}(\mathbf{r}_{n,m}) \nonumber \\
    = & \widetilde{\mathbf{h}}(\mathbf{s}) \mathbf{\Phi}_{\mathrm{R}} \mathbf{H} \mathbf{\Phi}_{\mathrm{T}} \mathbf{W}_t \mathbf{\Theta}_{t+1}^{-1}.
\end{align}
Based on the expression in \eqref{expression_g} and the transformation \eqref{GL_Tx}, $\mathbf{G}_{t+1}$ can be calculated by 
\begin{align}
    \mathbf{G}_{t+1} 
    = & \sum_{n=1}^M \sum_{m=1}^M \frac{\omega_n \omega_m A_{\mathrm{T}}}{4} \mathbf{g}^H_{t+1}(\mathbf{s}_{n,m}) \mathbf{g}_{t+1}(\mathbf{s}_{n,m}) \nonumber \\
    = &\mathbf{\Theta}_{t+1}^{-1} \mathbf{W}_t^H \mathbf{\Phi}_{\mathrm{T}} \mathbf{H}^H \mathbf{\Phi}_{\mathrm{R}} \mathbf{H} \mathbf{\Phi}_{\mathrm{T}} \mathbf{H}^H \mathbf{\Phi}_{\mathrm{R}} \mathbf{H} \mathbf{\Phi}_{\mathrm{T}} \mathbf{W}_t \mathbf{\Theta}_{t+1}^{-1},
\end{align}
Based on the expression in \eqref{expression_V} and the transformation \eqref{GL_Tx}, $\mathbf{V}_{t+1}$ can be calculated by 
\begin{align}
    \mathbf{V}_{t+1} 
    = &\sum_{n=1}^M \sum_{m=1}^M \frac{\omega_n \omega_m A_{\mathrm{R}}}{4} \mathbf{v}_{t+1}^H(\mathbf{r}_{n,m}) \mathbf{v}_{t+1}(\mathbf{r}_{n,m}) \nonumber \\
    = & \mathbf{\Theta}_{t+1}^{-1}  \mathbf{W}_t^H \mathbf{\Phi}_{\mathrm{T}} \mathbf{H}^H \mathbf{\Phi}_{\mathrm{R}} \mathbf{H} \mathbf{\Phi}_{\mathrm{T}} \mathbf{W}_t \mathbf{\Theta}_{t+1}^{-1}.
\end{align}
Therefore, based on \eqref{optimal_w_function}, $\mathbf{w}_{t+1}(\mathbf{s})$ can be calculated by 
\begin{align} \label{w_t_s}
    \mathbf{w}_{t+1}(\mathbf{s}) = & \mathbf{g}_{t+1}(\mathbf{s}) \mathbf{U}_{t+1} \left(\frac{1}{\varepsilon_{t+1}} \mathbf{I}_N + \mathbf{G}_{t+1} \mathbf{U}_{t+1} \right)^{-1} \nonumber \\
    = &\widetilde{\mathbf{h}}(\mathbf{s}) \mathbf{\Phi}_{\mathrm{R}} \mathbf{H} \mathbf{\Phi}_{\mathrm{T}} \mathbf{W}_t \mathbf{\Theta}_{t+1}^{-1} \mathbf{U}_{t+1} \mathbf{\Omega}_{t+1}^{-1},
\end{align}
where $\mathbf{\Omega}_{t+1} =  \frac{1}{\varepsilon_{t+1}} \mathbf{I}_N + \mathbf{G}_{t+1} \mathbf{U}_{t+1}$. 
Furthermore, the expression of $\mathbf{W}_{t+1}$ can be obtained as follows: 
\begin{align} \label{expression_Wt}
    \mathbf{W}_{t+1} &= \big[ \mathbf{w}_{t+1}^T(\mathbf{s}_{1,1}),\dots,\mathbf{w}_{t+1}^T(\mathbf{s}_{M,M})\big]^T \nonumber \\
    & = \mathbf{H}^H \mathbf{\Phi}_{\mathrm{R}} \mathbf{H} \mathbf{\Phi}_{\mathrm{T}} \mathbf{W}_t \mathbf{\Theta}_{t+1}^{-1} \mathbf{U}_{t+1} \mathbf{\Omega}_{t+1}^{-1}.
\end{align}

\subsubsection{Overall Implementation}
From the above results, it is evident that \textbf{Algorithm~\ref{alg:SU_WMMSE}} can be implemented by iteratively updating $\mathbf{W}_t$ according to \eqref{expression_Wt}. In each iteration, the objective value can be computed as  
\begin{equation}
    R_t = \log \det \left( \mathbf{I}_N + \frac{1}{\widetilde{\sigma}^2_{t}} \mathbf{Q}_t \right),
\end{equation}  
where $\mathbf{Q}_t$ is calculated as given in \eqref{expression_Q_t}.  
Thus, the overall matrix-based implementation of \textbf{Algorithm~\ref{alg:SU_WMMSE}} is summarized in \textbf{Algorithm~\ref{alg:matrix_WMMSE}}. This algorithm can be extended to solve beamforming problems with more general correlated power constraints, as discussed in Appendix \ref{appendix_extension}. The computational complexity of the algorithm arises primarily from matrix inversion and multiplication operations. \textcolor{black}{Therefore, given the dimensions of the involved matrices, the worst-case complexity is $O(N^3 + M_s^3 + NM_s^2 + M_sN^2)$.}

\begin{algorithm}[tb]
    \caption{Matrix-based WMMSE Algorithm for CAPA-MIMO Achievable Rate Maximization}
    \label{alg:matrix_WMMSE}
    \begin{algorithmic}[1]
        \STATE{initialize $t=0$ and $\mathbf{W}_0$}
        \REPEAT
            \STATE{update $\mathbf{W}_{t+1}$ as \eqref{expression_Wt} for given $\mathbf{W}_t$}
        \UNTIL{the fractional increase of the $R_t$ falls below a predefined threshold}
        \STATE{calculate $\mathbf{w}(\mathbf{s})$ according \eqref{w_t_s} using the converged $\mathbf{W}$ }
        \STATE{scale $\mathbf{w}(\mathbf{s})$ according to \eqref{scale_w}}
    \end{algorithmic}
\end{algorithm}

\subsection{\textcolor{black}{Number of Data Streams}}

\textcolor{black}{Unlike the eigendecomposition-based approach, which inherently determines the optimal number of data streams to maximize the multiplexing gain and achievable rate through water-filling, the WMMSE algorithm requires the number of data streams to be specified \emph{a priori}. A sensible choice is to set this number equal to the channel's spatial DoF.}

\textcolor{black}{In an NLoS environment with isotropic scattering, the DoF is given by
\begin{equation}
    \mathrm{DoF} = \frac{4\min\{A_{\mathrm{T}}, A_{\mathrm{R}}\}
    }{\lambda^2}.
\end{equation}
For LoS channels, extensive work (e.g.\cite{miller2000communicating, dardari2020communicating, 9896943}) has characterized their DoFs. However, most existing results are asymptotic. A well-known approximation is
\begin{equation} \label{eq_DoF}
    \mathrm{DoF} \approx \frac{A_{\mathrm{T}} A_{\mathrm{R}}}{ (\lambda D)^2},
\end{equation}
where $D$ is the transmitter-receiver separation. While such expressions capture the general scaling behavior, they apply strictly in the asymptotic regime and may not accurately predict DoF in more practical, finite-sized systems.}

\begin{figure}[t!]
    \centering
    \includegraphics[width=0.49\textwidth]{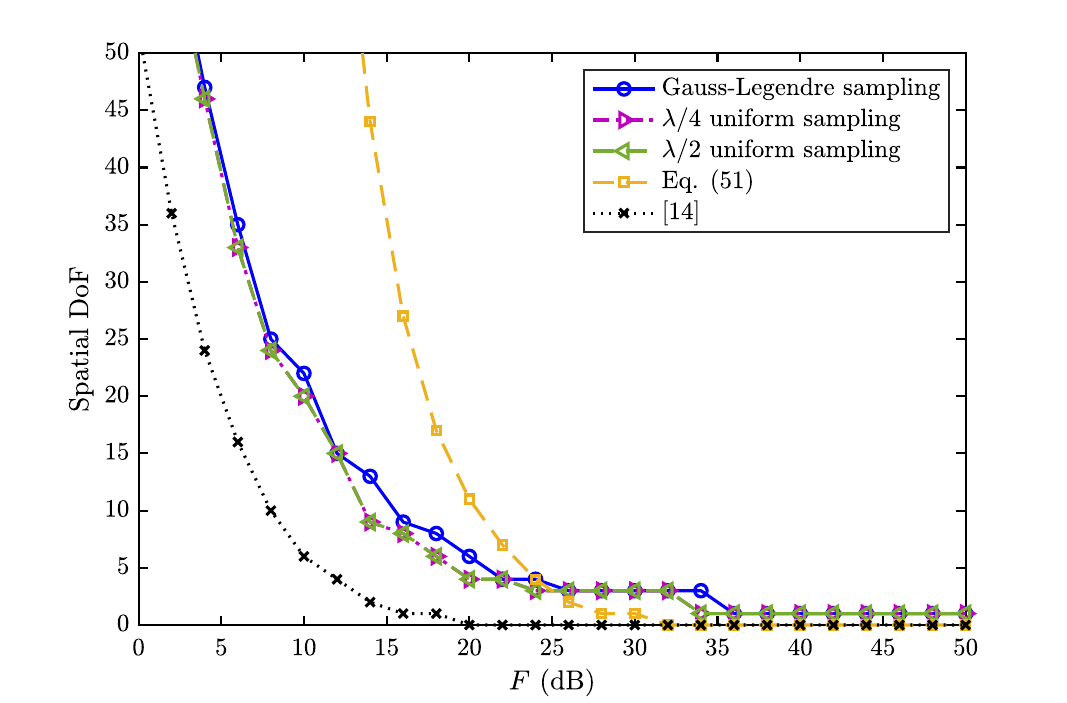}
    \caption{\textcolor{black}{Spatial DoF versus $F = D^2/A_{\mathrm{R}}$. }}
    \label{fig_DoF}
\end{figure}

\textcolor{black}{To address this issue, we propose numerically computing the DoF based on the singular values of the channel matrix $\mathbf{H}$ defined in \eqref{GL_channel}, which is constructed by sampling both the Tx- and Rx-CAPAs using the Gauss-Legendre quadrature rule. The resulting matrix has low dimensionality, making it well-suited for efficient DoF evaluation. In Fig.~\eqref{fig_DoF}, we compare the spatial DoF characterized using Gauss-Legendre sampling ($M = 8$) with other approaches for a system with $f = 5$ GHz and $A_{\mathrm{T}} = A_{\mathrm{R}} = 0.25 \text{ m}^2$. The DoF is determined by counting the number of singular values within 10 dB of the largest one. The proposed method achieves performance comparable to that of uniform sampling, which typically requires a much larger number of samples. Moreover, it provides better accuracy than the closed-form solution in \eqref{eq_DoF} and the approximation in \cite{dardari2020communicating}. }

\section{Comparison with Fourier-based Approach} \label{sec:Fourier}

In this section, we compare the proposed WMMSE algorithm with the existing Fourier-based approach for CAPA-MIMO beamforming. The Fourier-based method leverages the band-limited property of signals in the wavenumber domain (i.e., the Fourier transform of the spatial domain), enabling continuous signals to be represented by a finite set of Fourier basis functions. For the considered two-dimensional Tx- and Rx-CAPAs, the continuous  beamformer and receiver can be expressed via Fourier series expansions as
\begin{align}
    \mathbf{w}(\mathbf{s}) & = \sum_{n=-\infty}^{+\infty} \sum_{m=-\infty}^{+\infty} \bm{w}_{n,m} \psi_{\mathrm{T}, n,m}(\mathbf{s}), \\
    \mathbf{v}(\mathbf{r}) & = \sum_{n=-\infty}^{+\infty} \sum_{m=-\infty}^{+\infty} \bm{v}_{n,m} \psi_{\mathrm{R},n,m}(\mathbf{r}),
\end{align}
where $\bm{w}_{n,m} \in \mathbb{C}^{1 \times N}$ and $\bm{v}_{n,m} \in \mathbb{C}^{1 \times N}$ denote the beamformer and receiver coefficients in the wavenumber domain, respectively. The Fourier basis functions $\psi_{\mathrm{T},n,m}(\mathbf{s})$ and $\psi_{\mathrm{R},n,m}(\mathbf{r})$ are defined as \cite{9906802, 9724113}
\begin{align}
    \psi_{\mathrm{T}, n,m}(\mathbf{s}) = & \frac{1}{\sqrt{A_{\mathrm{T}}}} e^{j \left( \frac{2\pi }{L_{\mathrm{T},x}} n s_x + \frac{2 \pi }{L_{\mathrm{T},y}} m s_y \right)}, \\
    \psi_{\mathrm{R}, n,m}(\mathbf{r}) = & \frac{1}{\sqrt{A_{\mathrm{R}}}} e^{j \left( \frac{2\pi }{L_{\mathrm{R},x}} n \hat{r}_x + \frac{2 \pi }{L_{\mathrm{R},y}} m \hat{r}_y \right)}.
\end{align}
The wavenumber-domain representations of the beamformer and receiver are obtained by
\begin{align} \label{expression_transmit_signal_WD}
    \bm{w}_{n,m} = & \int_{\mathcal{S}_{\mathrm{T}}} \mathbf{w}(\mathbf{s}) \psi_{\mathrm{T}, n,m}^H(\mathbf{s}) d \mathbf{s}, \\
    \bm{v}_{n,m} = & \int_{\mathcal{S}_{\mathrm{R}}} \mathbf{v}(\mathbf{r}) \psi_{\mathrm{R}, n,m}^H(\mathbf{r}) d \mathbf{r}.
\end{align}
It is proved that spatial-domain signals are band-limited in the wavenumber domain, leading to the following approximations~\cite{9906802, 9724113}
\begin{align} \label{WD_approx}
    \mathbf{w}(\mathbf{s}) \approx & \sum_{(n,m) \in \mathcal{D}_{\mathrm{T}}} \bm{w}_{n,m} \psi_{\mathrm{T}, n,m}(\mathbf{s}), \\
    \mathbf{v}(\mathbf{r}) \approx & \sum_{(n,m) \in \mathcal{D}_{\mathrm{R}}}  \bm{v}_{n,m} \psi_{\mathrm{R}, n,m}(\mathbf{r}),
\end{align}
where the index sets are defined as 
\begin{align}
    \mathcal{D}_{\mathrm{T}} = &\left\{ (n,m) \in \mathbb{Z}^2 : |n| \le \left\lceil \frac{L_{\mathrm{T},x}}{\lambda} \right\rceil, |m| \le \left\lceil \frac{L_{\mathrm{T},y}}{\lambda} \right\rceil \right\}, \\
    \mathcal{D}_{\mathrm{R}} = &\left\{ (n,m) \in \mathbb{Z}^2 : |n| \le \left\lceil \frac{L_{\mathrm{R},x}}{\lambda} \right\rceil, |m| \le \left\lceil \frac{L_{\mathrm{R},y}}{\lambda} \right\rceil \right\}.
\end{align}
Define $M_{\mathrm{T},x} = \left\lceil \frac{L_{\mathrm{T},x}}{\lambda} \right\rceil$, $M_{\mathrm{T},y} = \left\lceil \frac{L_{\mathrm{T},y}}{\lambda} \right\rceil$, $M_{\mathrm{R},x} = \left\lceil \frac{L_{\mathrm{R},x}}{\lambda} \right\rceil$, and $M_{\mathrm{R},y} = \left\lceil \frac{L_{\mathrm{R},y}}{\lambda} \right\rceil$. Thus, the total number of terms in the approximated Fourier series for the beamformer and receiver are  $\widetilde{M}_{\mathrm{T}} = (2M_{\mathrm{T}, x}+1) (2M_{\mathrm{T},y}+1)$ and $\widetilde{M}_{\mathrm{R}} = (2M_{\mathrm{R}, x}+1) (2M_{\mathrm{R},y}+1)$, respectively.

Based on the approximation discussed above, the estimated signal at the RX-CAPA after applying the receiver $\mathbf{v}(\mathbf{r})$ can be expressed as 
\begin{align}
    \hat{\mathbf{y}} = & \int_{\mathcal{S}_{\mathrm{R}}} \mathbf{v}^H(\mathbf{r}) y(\mathbf{r}) d \mathbf{r} \nonumber \\
    \approx & \sum_{(n,m)\in \mathcal{D}_{\mathrm{R}}} \sum_{(n',m')\in \mathcal{D}_{\mathrm{T}}}  \bm{v}_{n,m}^H H^w_{n,m,n',m'} \bm{w}_{n',m'} \mathbf{c} + \bm{z},
\end{align} 
where $\bm{z} \sim \mathcal{CN} \Big( \mathbf{0}, \sigma^2 \sum_{(n,m)\in \mathcal{D}_{\mathrm{R}}}  \bm{v}_{n,m}^H \bm{v}_{n,m}  \Big)$ and the wavenumber-domain channel  $H^w_{n,m,n',m'}$ is given by
\begin{align} \label{WD_channel_each}
    H^w_{n,m,n',m'} = \int_{\mathcal{S}_{\mathrm{R}}} \int_{\mathcal{S}_{\mathrm{T}}} \psi_{\mathrm{R}, n,m}^H(\mathbf{r}) h(\mathbf{r}, \mathbf{s}) \psi_{\mathrm{T}, n',m'}(\mathbf{s}) d \mathbf{s} d \mathbf{r}.
\end{align}
To simplify notation, we define the following matrices:
\begin{align} \label{WD_channel_matrix}
    \bm{H}_w = &\begin{bmatrix}
        H_{1,1,1,1}^w & \cdots & H_{1,1,M_{\mathrm{T},x},M_{\mathrm{T},y}}^w \\
        \vdots & \ddots  & \vdots\\
        H_{M_{\mathrm{R},x},M_{\mathrm{R},y},1,1}^w &  \cdots  & H_{M_{\mathrm{R},x},M_{\mathrm{R},y},M_{\mathrm{T},x},M_{\mathrm{T},y}}^w
    \end{bmatrix} \nonumber \\
    & \hspace{5.1cm} \in \mathbb{C}^{\widetilde{M}_{\mathrm{R}} \times \widetilde{M}_{\mathrm{T}}}  \\
    \bm{W} = &\left[\bm{w}_{1,1},\dots,\bm{w}_{M_{\mathrm{T},x}, M_{\mathrm{T},y}}\right] \in \mathbb{C}^{\widetilde{M}_{\mathrm{T}} \times N},  \\
    \bm{V} = &\left[\bm{v}_{1,1},\dots,\bm{v}_{M_{\mathrm{R},x}, M_{\mathrm{R},y}}\right] \in \mathbb{C}^{\widetilde{M}_{\mathrm{R}} \times N}.
\end{align}
Consequently, the estimated signal simplifies to
\begin{equation}
    \hat{\mathbf{y}} = \bm{V}^H \bm{H}_w \bm{W} \mathbf{c} + \bm{z},
\end{equation}
which resembles the conventional discrete MIMO system model. Thus, the achievable rate can be approximated as
\begin{equation}
    R \approx \log \det \left( \mathbf{I}_N + \frac{1}{\sigma^2} \bm{W}^H \bm{H}_w^H \bm{H}_w \bm{W}  \right).
\end{equation}
Similarly, the transmit power can be approximated by 
\begin{align}
    \int_{\mathcal{S}_{\mathrm{T}}} \|\mathbf{w}(\mathbf{s})\|^2 d \mathbf{s} \approx \mathrm{Tr}\left( \bm{W} \bm{W}^H  \right),
\end{align}
resulting in the following achievable rate maximization problem in the wavenumber domain:
\begin{subequations} \label{WD_problem}
    \begin{align}
        \max_{\bm{W}} \quad &\log \det \left( \mathbf{I}_N + \frac{1}{\sigma^2} \bm{W}^H \bm{H}_w^H \bm{H}_w \bm{W} \right) \\
        \mathrm{s.t.} \quad & \mathrm{Tr}\left( \bm{W} \bm{W}^H  \right) \le P_{\mathrm{T}}.
    \end{align}
\end{subequations} 
This formulation aligns exactly with the standard discrete MIMO beamforming problem. The optimal solution for $\bm{W}$ can be found through the SVD of the matrix $\bm{H}_w$ and the water-filling power allocation \cite{1203154}. 

\textcolor{black}{Although both the proposed WMMSE approach and the Fourier-based approach involve approximations, the nature of these approximations is fundamentally different. Specifically, the approximation in the proposed WMMSE approach arises solely from the use of Gauss-Legendre quadrature \eqref{GL_quadrature}, which introduces only numerical integration error. This error diminishes rapidly as the number of samples increases, thanks to the mathematically optimal selection of sampling points and weights in Gauss-Legendre quadrature. In contrast, the approximation in the Fourier-based approach stems primarily from \eqref{WD_approx}, which introduces a modeling error due to the truncation of the Fourier series representation.}

The computational complexity of the Fourier-based approach primarily arises from three components. The first component is the calculation of the wavenumber-domain channel matrix $\bm{H}_w$ with entries defined in \eqref{WD_channel_each}. Each entry can be calculated using the Gauss-Legendre quadrature and the transformation in \eqref{GL_Tx} and \eqref{GL_Rx}, given by
\begin{equation}
    H^w_{n,m,n',m'} = \boldsymbol{\psi}_{\mathrm{R},n,m}^H \mathbf{\Phi}_{\mathrm{R}} \mathbf{H} \mathbf{\Phi}_{\mathrm{T}} \boldsymbol{\psi}_{\mathrm{T},n',m'},
\end{equation}
where 
\begin{align}
    \boldsymbol{\psi}_{\mathrm{T},n,m} &= \left[ \psi_{\mathrm{T},n,m}(\mathbf{s}_{1,1}),\dots, \psi_{\mathrm{T},n,m}(\mathbf{s}_{M,M}) \right]^T, \\
    \boldsymbol{\psi}_{\mathrm{R},n,m} &= \left[ \psi_{\mathrm{R},n,m}(\mathbf{r}_{1,1}),\dots, \psi_{\mathrm{R},n,m}(\mathbf{r}_{M,M}) \right]^T.
\end{align}
\textcolor{black}{Thus, the complexity for computing $\bm{H}_w$ is $O(\widetilde{M}_{\mathrm{R}} \widetilde{M}_{\mathrm{T}} M_s^2) = O(A_{\mathrm{R}} A_{\mathrm{T}} M_s^2 / \lambda^4)$. The second component is the SVD of $\bm{H}_w$ with complexity $\mathcal{O}(\widetilde{M}_{\mathrm{R}} \widetilde{M}_{\mathrm{T}} \max\{\widetilde{M}_{\mathrm{R}}, \widetilde{M}_{\mathrm{T}}\}) = O(A_{\mathrm{R}} A_{\mathrm{T}} \max\{A_{\mathrm{R}}, A_{\mathrm{T}}\}/\lambda^6)$. The third component is the power allocation using water-filling, which has a complexity of $O(N)$ \cite{1381759}. It can be observed that the computational complexity of the Fourier-based approach is primarily determined by $\widetilde{M}_{\mathrm{T}}$ and $\widetilde{M}_{\mathrm{R}}$, which increase with both aperture size and carrier frequency and can be extremely large. For example, for a Tx-CAPA with $L_{\mathrm{T},x} = L_{\mathrm{T},y} = 0.5$ m, the value of $\widetilde{M}_{\mathrm{T}}$ reaches $81$ at $2.4$ GHz, $729$ at $7.8$ GHz, and $2601$ at $15$ GHz. Consequently, although the Fourier-based approach effectively converts the CAPA-MIMO problem into a discrete MIMO problem, it results in unaffordable computational complexity compared to the proposed WMMSE approach.}

\section{Numerical Results} \label{sec:results}


This section presents numerical results demonstrating the effectiveness of the proposed WMMSE algorithm. \textcolor{black}{Unless otherwise specified, all simulations are conducted using the following setup \cite{9906802, zhang2023pattern}.} The areas of the Tx- and Rx-CAPAs are both set as $A_{\mathrm{T}} = A_{\mathrm{R}} = 0.25\,\text{m}^2$, with edge lengths $L_{\mathrm{T},x} = L_{\mathrm{T},y} = \sqrt{A_{\mathrm{T}}}$ and $L_{\mathrm{R},x} = L_{\mathrm{R},y} = \sqrt{A_{\mathrm{R}}}$. The position and orientation of the Rx-CAPA relative to the Tx-CAPA are defined by setting $\mathbf{r}_o = [0, 0, 10\,\text{m}]^T$ and $\alpha = \beta = \phi = 0$, which implies that the Tx- and Rx-CAPAs are parallel to each other and their polarization directions are matched for maximum wave reception. The signal frequency is set to $f = 2.4\,\text{GHz}$, and the free-space impedance is $\eta = 120\pi\,\Omega$. The transmit power and noise power are specified as $P_{\mathrm{T}} = 100\,\text{mA}^2$ and $\sigma^2 = 5.6 \times 10^{-3}\,\text{V}^2/\text{m}^2$, respectively. Finally, the number of Gauss-Legendre quadrature samples is set to $M = 10$. 

\subsection{Benchmarks}
The following benchmarks are considered in the simulation.

\subsubsection{Fourier-SVD}
This benchmark refers to the Fourier-based approach presented in Section \ref{sec:Fourier}, where the continuous beamformer and receiver are approximated using finite Fourier series and the final problem is solved using SVD of the wavenumber-domain channel. \textcolor{black}{For both the proposed WMMSE algorithm and the Fourier-SVD benchmark, the number of data streams is set to $N = \min\{ \widetilde{M}_{\mathrm{R}}, \widetilde{M}_{\mathrm{T}} \}$, unless stated otherwise, to ensure the maximization of the multiplexing gain.}

\subsubsection{SPDA}
In this benchmark, both the transmitter and receiver are assumed to utilize SPDAs of identical size to their CAPA counterparts. The corresponding signal model is derived following the methods presented in \cite{zhang2023pattern} and \cite{9906802}.  Specifically, the Tx-SPDA and Rx-SPDA are each composed of discrete antennas with an effective aperture of $A_d = \frac{\lambda^2}{4\pi}$ and spacing of $d = \frac{\lambda}{2}$. Consequently, the number of antennas is $\overline{M}_{\mathrm{T}} = \overline{M}_{\mathrm{T}, x} \times \overline{M}_{\mathrm{T}, y}$ for the Tx-SPDA and $\overline{M}_{\mathrm{R}} = \overline{M}_{\mathrm{R}, x} \times \overline{M}_{\mathrm{R}, y}$ for the Rx-SPDA, where $\overline{M}_{\mathrm{T},x} = \lceil \frac{L_{\mathrm{T},x}}{d} \rceil$, $\overline{M}_{\mathrm{T},y} = \lceil \frac{L_{\mathrm{T},y}}{d} \rceil$, $\overline{M}_{\mathrm{R},x} = \lceil \frac{L_{\mathrm{R},x}}{d} \rceil$ and $\overline{M}_{\mathrm{R},y} = \lceil \frac{L_{\mathrm{R},y}}{d} \rceil$. The position of the $(n,m)$-th antenna at the Tx-SPDA is given by 
\begin{align}
  \bar{\mathbf{s}}_{n,m} = \left[ (n-1)d - \frac{L_{\mathrm{T}, x}}{2}, (m-1)d - \frac{L_{\mathrm{T}, y}}{2}, 0 \right]^T.
\end{align}
Similarly, the location of the $(n,m)$-th antenna at the Rx-SPDA is given by
\begin{align}
  \bar{\mathbf{r}}_{n,m} &= \mathbf{R}_z(\alpha) \mathbf{R}_y(\beta) \mathbf{R}_x(\phi)  \hat{\bar{\mathbf{r}}}_{n,m} + \mathbf{r}_o, \\
  \hat{\bar{\mathbf{r}}}_{n,m} &= \left[ (n-1)d - \frac{L_{\mathrm{R}, x}}{2}, (m-1)d - \frac{L_{\mathrm{R}, y}}{2}, 0 \right]^T.
\end{align}  
Let $\mathcal{S}_{\mathrm{T},n,m}$ and $\mathcal{S}_{\mathrm{R},n,m}$ denote the surface of the $(n,m)$-th antennas at the Tx-SPDA and Rx-SPDA, respectively, satisfying $\left|\mathcal{S}_{\mathrm{T},n,m}\right| = \left|\mathcal{S}_{\mathrm{R},n,m}\right| = A_d$. The beamformer and receiver can be expressed as 
\begin{align}
  \overline{\mathbf{w}}(\mathbf{s}) & = \frac{1}{\sqrt{A_d}} \sum_{(n,m) \in \mathcal{E}_{\mathrm{T}}} \overline{\bm{w}}_{n,m} \mathrm{rect}(\mathbf{s} \in \mathcal{S}_{\mathrm{T}, n,m}), \\
  \overline{\mathbf{v}}(\mathbf{r}) & = \frac{1}{\sqrt{A_d}} \sum_{(n,m) \in \mathcal{E}_{\mathrm{R}}} \overline{\bm{v}}_{n,m} \mathrm{rect}(\mathbf{r} \in \mathcal{S}_{\mathrm{R},n,m}),
\end{align}
where $\overline{\bm{w}}_{n,m} \in \mathbb{C}^{1 \times N}$ and $\overline{\bm{v}}_{n,m} \in \mathbb{C}^{1 \times N}$ are the discrete beamformer and receiver, respectively. The index sets are given by  $\mathcal{E}_{\mathrm{T}} = \{ (n,m) \in \mathbb{Z}^2 : 1 \le n \le \overline{M}_{\mathrm{T}, x}, 1 \le m \le \overline{M}_{\mathrm{T}, y} \}$ and $\mathcal{E}_{\mathrm{R}} = \{ (n,m) \in \mathbb{Z}^2 : 1 \le n \le \overline{M}_{\mathrm{R}, x}, 1 \le m \le \overline{M}_{\mathrm{R}, y}  \}$. The SPDA signal model can thus be obtained following the procedure described in Section~\ref{sec:Fourier}. \textcolor{black}{Specifically, referring to \eqref{WD_channel_each}, the channel between the $(n',m')$-th transmit antenna and the $(n,m)$-th receive antenna is given by 
\begin{align}
  \overline{H}_{n,m,n',m'} = &\int_{\mathcal{S}_{\mathrm{R}, n, m}} \int_{\mathcal{S}_{\mathrm{T}, n', m'}} \frac{1}{A_d} h(\mathbf{r}, \mathbf{s}) d \mathbf{r} d \mathbf{s} \nonumber \\
  \approx & \left|\mathcal{S}_{\mathrm{R}, n, m}\right| \left|\mathcal{S}_{\mathrm{T}, n', m'}\right| \frac{1}{A_d} h(\bar{\mathbf{r}}_{n,m}, \bar{\mathbf{s}}_{n',m'})  \nonumber \\
  = & A_d h(\bar{\mathbf{r}}_{n,m}, \bar{\mathbf{s}}_{n',m'}),
\end{align}   
where the approximation is made under the assumption that $A_d = \frac{\lambda^2}{4\pi}$ is sufficiently small.} Consequently, the achievable rate is given by 
\begin{equation}
  \overline{R} = \log \det \left( \mathbf{I}_{N} + \frac{1}{\sigma^2} \overline{\bm{W}}^H \overline{\bm{H}}^H \overline{\bm{H}} \overline{\bm{W}}  \right),
\end{equation}
where $\overline{\bm{W}} = [ \overline{\bm{w}}_{1,1}^T,\dots,\overline{\bm{w}}_{\overline{M}_{\mathrm{T},x},\overline{M}_{\mathrm{T},y}}^T ]^T$ and $\overline{\bm{H}}$ is defined similar to \eqref{WD_channel_matrix}. The optimal solution to the corresponding rate maximization problem can be obtained through SVD and water-filling power allocation. For the SPDA benchmark, the number of data streams is set to $N = \min\{ \overline{M}_{\mathrm{R}}, \overline{M}_{\mathrm{T}} \}$.


\begin{table*}
  \centering
  \caption{Comparison of Average CPU Time for Different Approaches with $N = 10$ and $M$ = 10.}
  \label{table_CPU_time}
  \begin{tabular}{c|cc|cc|cc}
    \toprule
    \multirow{2}{*}{Frequency} &
      \multicolumn{2}{c|}{$A_{\mathrm{T}} = A_{\mathrm{R}} = 0.2$ m$^2$} &
      \multicolumn{2}{c|}{$A_{\mathrm{T}} = A_{\mathrm{R}} = 0.3$ m$^2$} &
      \multicolumn{2}{c}{$A_{\mathrm{T}} = A_{\mathrm{R}} = 0.4$ m$^2$} \\
      & WMMSE & Fourier-SVD & WMMSE & Fourier-SVD & WMMSE & Fourier-SVD \\
      \midrule
    2.4 GHz & 0.246 s & 0.615 s & 0.244 s & 1.277 s & 0.257 s & 2.387 s  \\
    5 GHz & 0.245 s & 7.029 s & 0.255 s & 16.076 s & 0.237 s & 22.828 s \\
    7.8 GHz & 0.239 s & 31.985 s & 0.247 s & 75.893 s & 0.243 s & 123.490 s \\
    \bottomrule
    \multicolumn{7}{l}{\makecell[l]{\textcolor{black}{Note that the CPU time of the proposed WMMSE algorithm can be significantly lower than} \\ \textcolor{black}{the results in this table, as it requires only $M < 5$ samples according to Fig. \ref{fig_GL_convergence}. In contrast, } \\ \textcolor{black}{the Fourier-SVD approach typically requires $M = 10 \sim 20$, resulting in much higher comput-} \\ \textcolor{black}{ational complexity.}}}
  \end{tabular}
  \vspace{-0.5cm}
\end{table*}

\subsection{Convergence and Complexity of the Proposed Algorithm}

\begin{figure}[t!]
  \centering
  \includegraphics[width=0.49\textwidth]{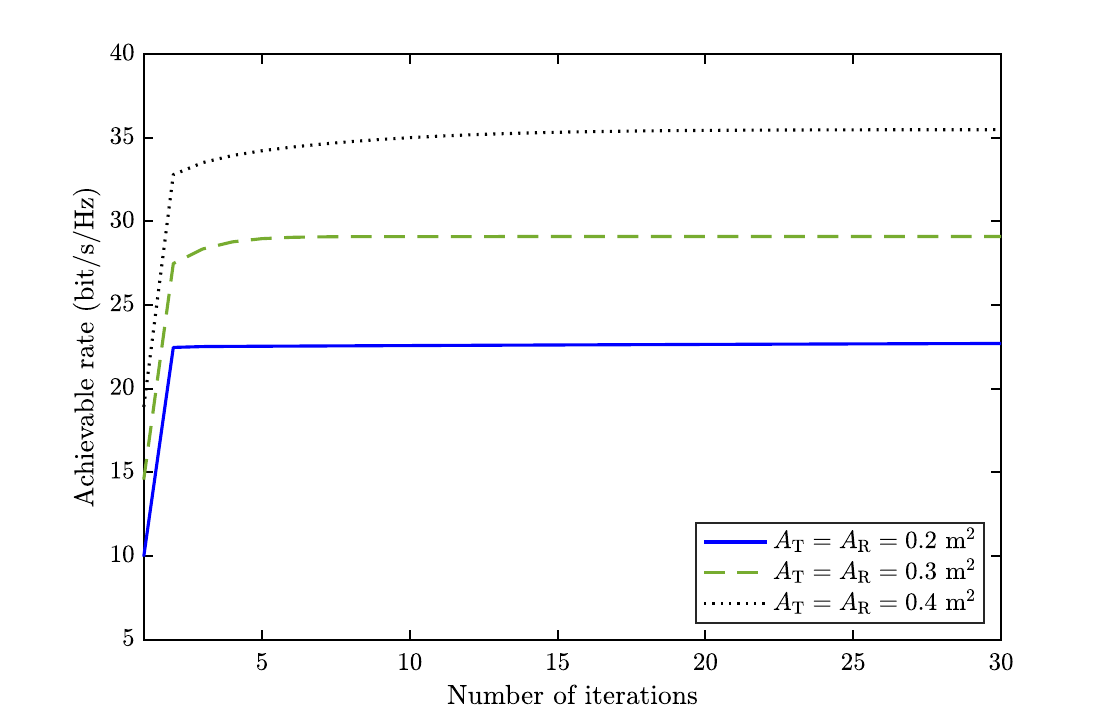}
  \caption{Convergence behavior of the proposed WMMSE algorithm.}
  \label{fig_convergence}

  \centering
  \includegraphics[width=0.49\textwidth]{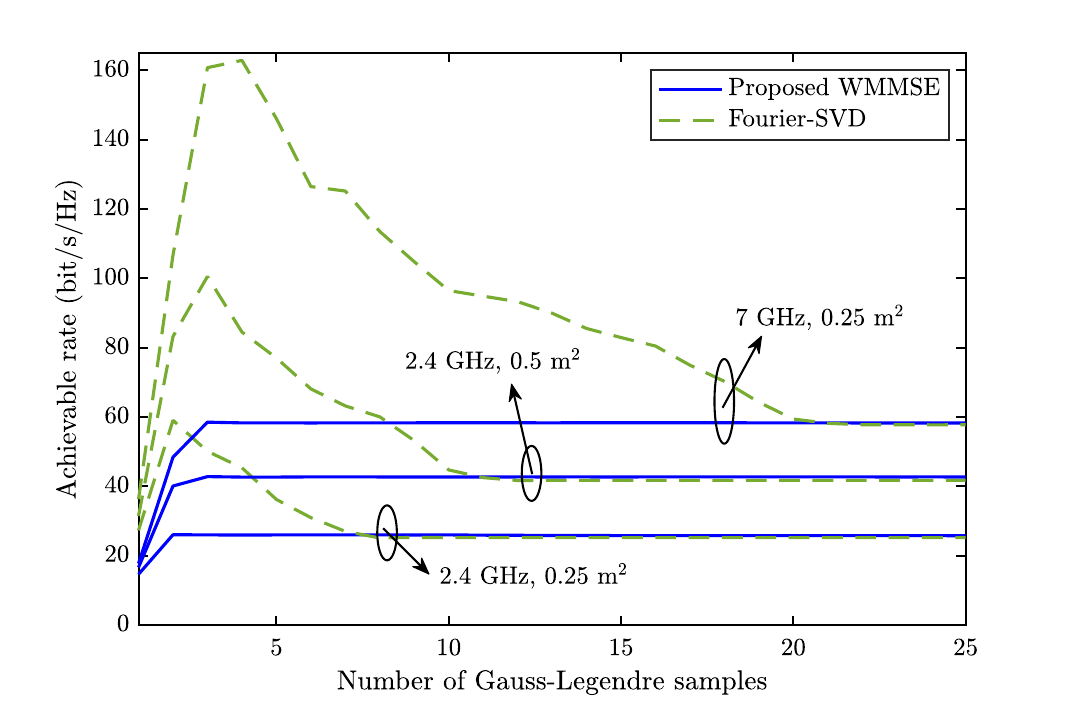}
  \caption{Convergence behavior of the achievable rate with the number Gauss-Legendre samples under different signal frequencies and aperture sizes. \textcolor{black}{Note that any result obtained before the convergence of Gauss-Legendre quadrature is not meaningful, as the numerical integration is still insufficiently accurate and cannot reliably reflect the true value. Therefore, only the results obtained after convergence should be considered valid.}}
  \label{fig_GL_convergence}
\end{figure}

Fig.~\ref{fig_convergence} demonstrates the rapid convergence of the proposed WMMSE algorithm with respect to the number of iterations under varying aperture sizes, thereby confirming its effectiveness. Furthermore, as illustrated in Fig.~\ref{fig_GL_convergence}, another convergence trend is observed as the number of Gauss-Legendre samples increases, owing to the enhanced accuracy in integral computations. \textcolor{black}{Notably, the proposed WMMSE algorithm requires only a few samples (fewer than $5$) to achieve high accuracy, whereas the Fourier-SVD approach demands significantly more, especially for high frequencies and large apertures. This disparity arises because the integrals in the wavenumber-domain channel matrix $\bm{H}_w$ may exhibit some non-desirable properties, such as insufficient smoothness, high-frequency oscillations, or rapidly increasing derivatives.} These observations indicate that the high computational complexity of the Fourier-SVD approach stems not only from the large number of wavenumber-domain samples but also from the extensive Gauss-Legendre samples required.

To further investigate computational complexity, Table~\ref{table_CPU_time} presents the CPU time consumed by different approaches, with parameters fixed at $N = 10$ and $M = 10$. \textcolor{black}{All experiments are conducted under a controlled software environment using a single-threaded MATLAB implementation on an Apple M3 silicon chip, and the number of iterations for the proposed WMMSE algorithm is fixed at $100$.} It is observed that the CPU time for the Fourier-SVD approach increases dramatically from $0.615$ s to $123.490$ s as frequency and array aperture grow, whereas the proposed WMMSE algorithm maintains a consistently low CPU time of approximately $0.250$ s. The real CPU time of the WMMSE algorithm can be even lower than $0.250$ seconds, as the number of iterations required for convergence is typically much less than $100$, as demonstrated in Fig.~\ref{fig_convergence}. \textcolor{black}{It is worth noting that the results in Table~\ref{table_CPU_time} aims to offer a baseline comparison of computational complexity between the proposed WMMSE algorithm and the Fourier-based approach under identical software conditions. This baseline does not fully reflect performance in practical implementations, especially on platforms like FPGAs or GPUs, which offer substantial architectural advantages and parallelization opportunities.}

\begin{figure}[t!]
  \centering
  \includegraphics[width=0.49\textwidth]{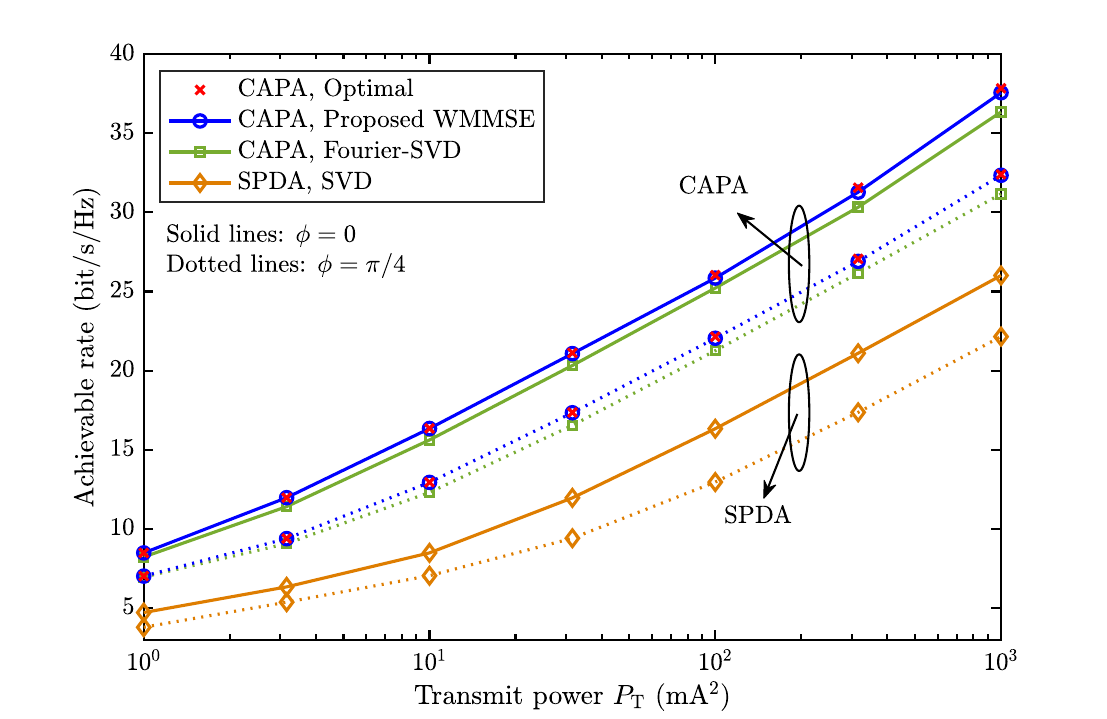}
  \caption{Achievable rate versus transmit power.}
  \label{fig_transmit_power}

  \centering
  \includegraphics[width=0.49\textwidth]{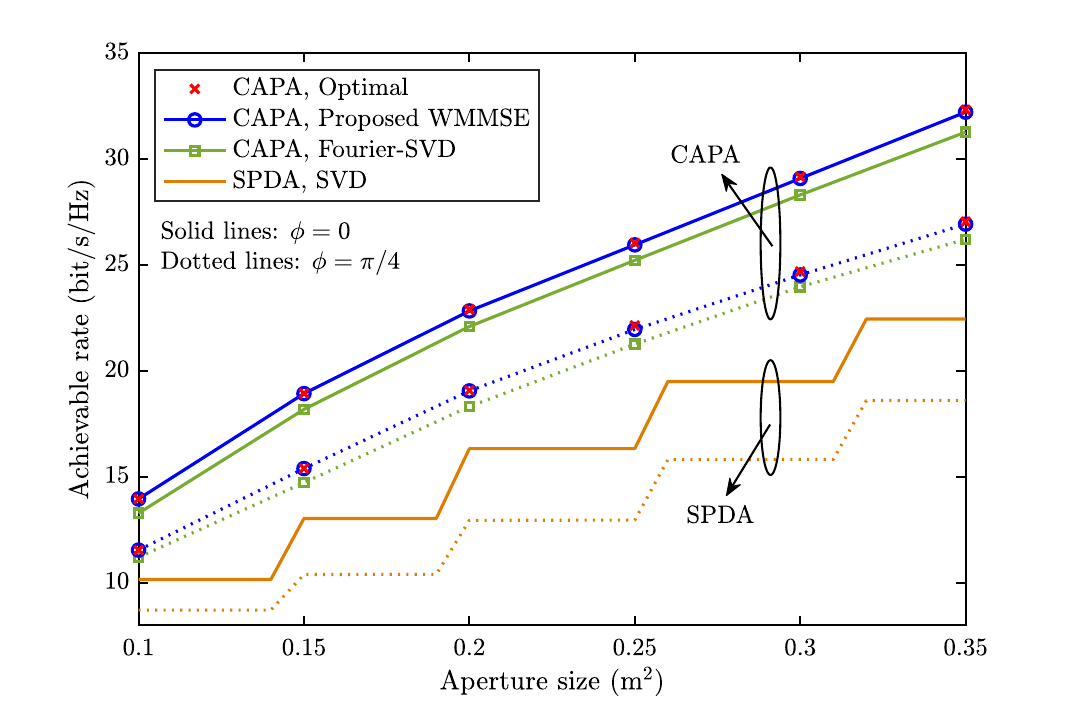}
  \caption{Achievable rate versus aperture size.}
  \label{fig_aperture_size}
\end{figure} 


\subsection{Achievable Rate Versus Different Parameters}

\begin{figure}[t!]
  \centering
  \includegraphics[width=0.49\textwidth]{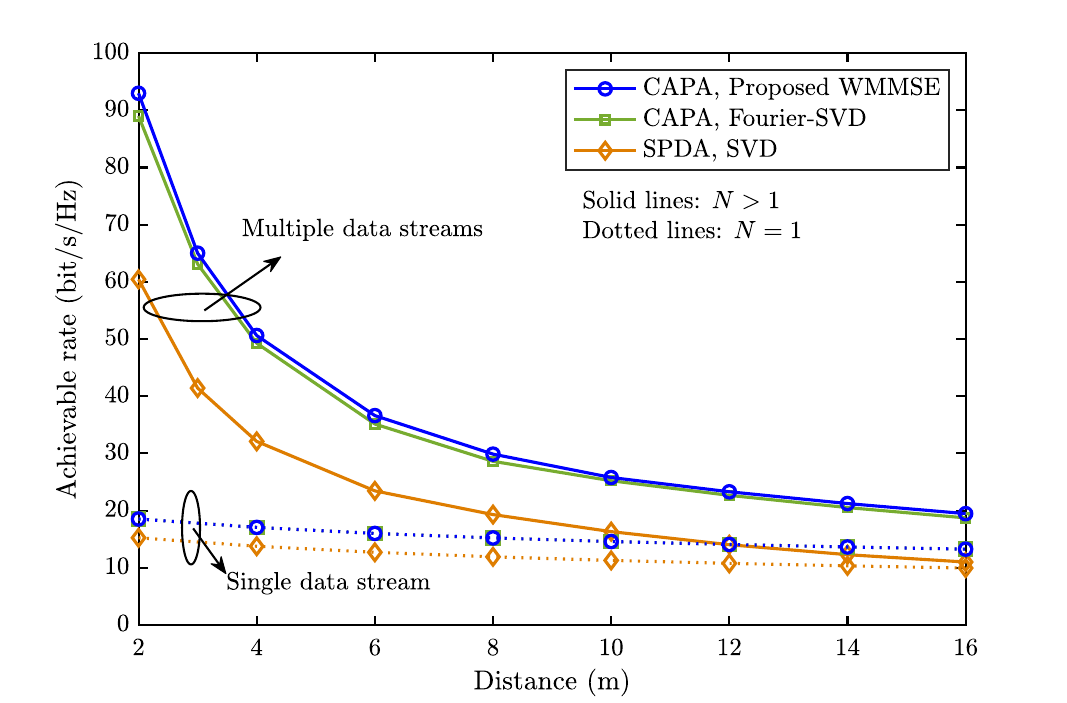}
  \caption{Achievable rate versus distance between Tx- and Rx-CAPAs.}
  \label{fig_distance}

  \centering
  \includegraphics[width=0.49\textwidth]{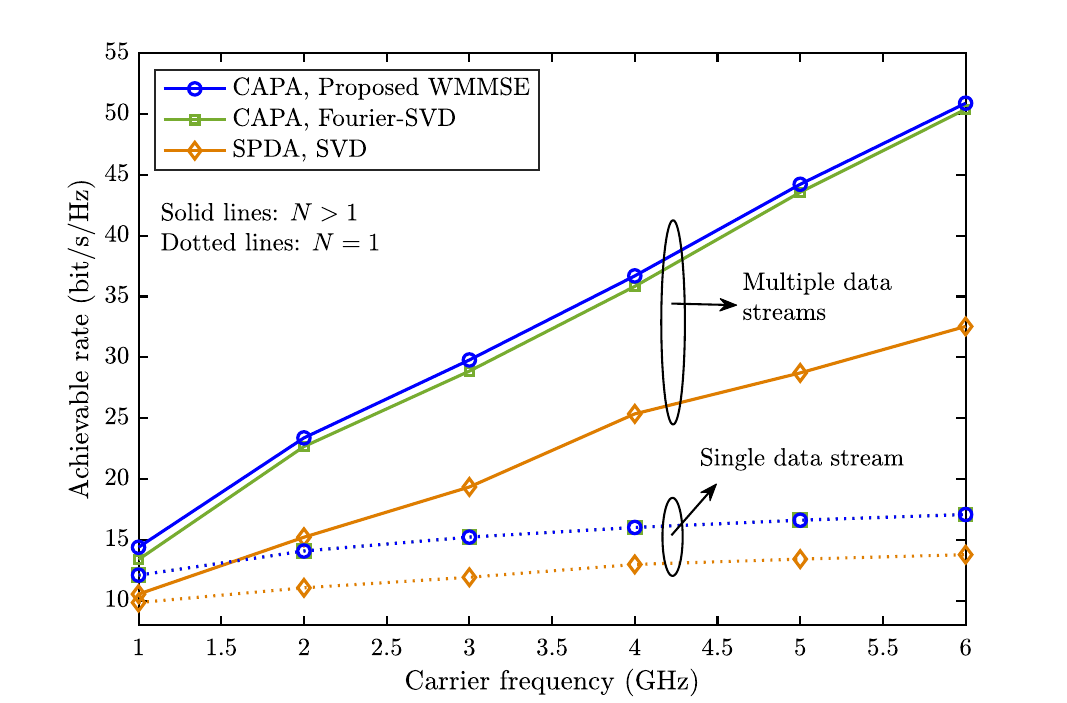}
  \caption{\textcolor{black}{Achievable rate versus carrier frequency.}}
  \label{fig_frequency}
\end{figure}


Fig.~\ref{fig_transmit_power} illustrates the increase in achievable rate with higher transmit power. \textcolor{black}{Here, we also the optimal approach as a baseline, which is obtained by densely sampling the apertures as given in \cite{dardari2020communicating}. As shown in the figure, the proposed WMMSE algorithm consistently achieves a near-optimal achievable rate, which is higher than the Fourier-SVD approach, while also exhibiting lower computational complexity.} This advantage stems from the ability of the proposed algorithm to effectively avoid discretization loss. The figure also highlights the significant performance gain of CAPA over SPDA, attributed to the enhanced spatial DoFs offered by CAPA. Additionally, a performance loss occurs when the rotation angle $\phi$, which defines the orientation of the Rx-CAPA with respect to the $x$-axis, changes from $0$ to $\pi/4$. This performance loss results from the reduced effective apertures and polarization mismatch between non-parallel CAPAs \cite{10220205}. Similar phenomenons are observed in Fig. \ref{fig_aperture_size}, which demonstrates the impact of array aperture. In this figure, we keep $A_{\mathrm{T}} = A_{\mathrm{R}}$. It is interesting to see that the achievable rate of SPDAs does not increase smoothly with aperture size. This is because a larger aperture does not always result in a higher number of antennas for SPDAs.

\begin{figure}[t!]
  \centering
  \includegraphics[width=0.49\textwidth]{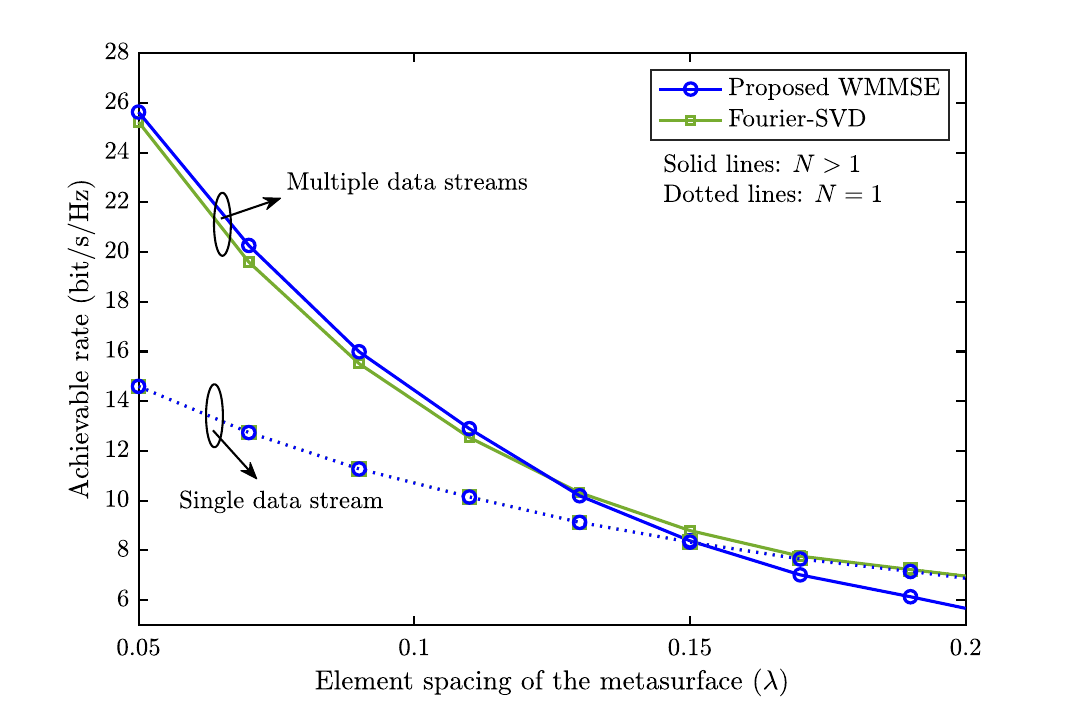}
  \caption{\textcolor{black}{Achievable rate versus element spacing of the metasurface.}}
  \label{fig_metasurface}
\end{figure}

\begin{figure*}[t!]
  \centering
  \begin{subfigure}[t]{0.25\textwidth}
    \includegraphics[width=1\textwidth]{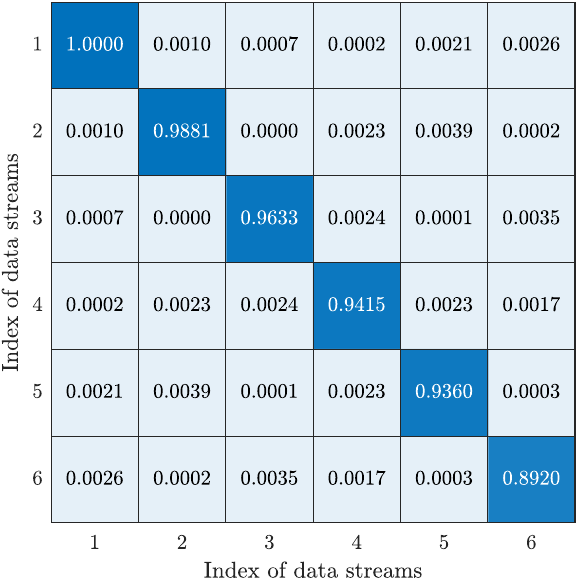}
    \caption{$N = 6$, $R = 42.08$ bit/s/Hz.}
  \end{subfigure}
  \hspace{1cm}
  \begin{subfigure}[t]{0.25\textwidth}
    \includegraphics[width=1\textwidth]{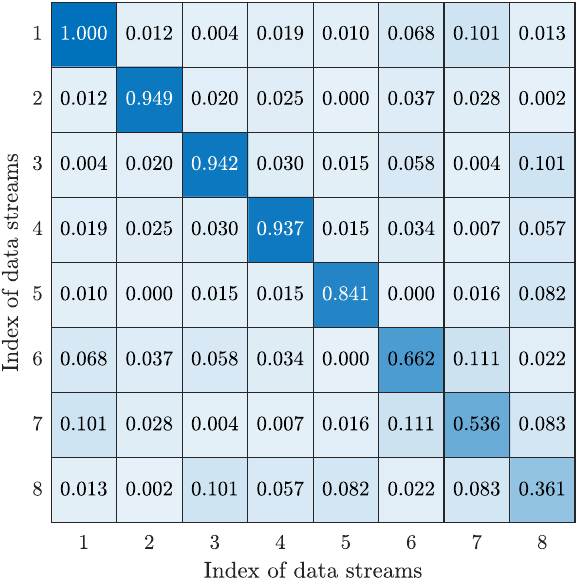}
    \caption{$N = 8$, $R = 42.14$ bit/s/Hz.}
  \end{subfigure}
  \hspace{1cm}
  \begin{subfigure}[t]{0.25\textwidth}
    \includegraphics[width=1\textwidth]{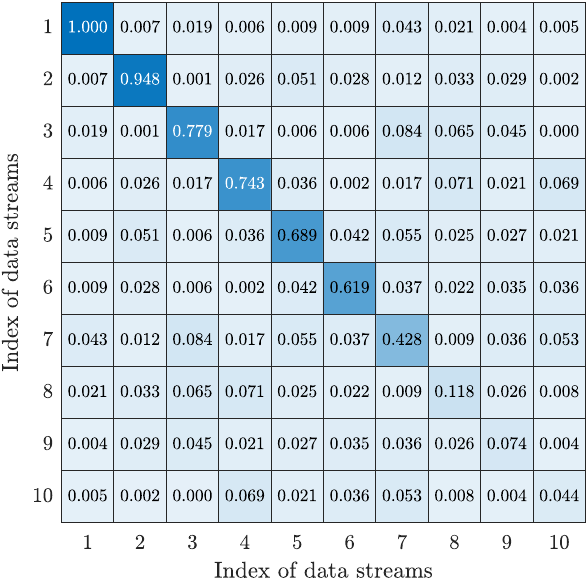}
    \caption{$N = 10$, $R = 42.26$ bit/s/Hz.}
  \end{subfigure}
  \caption{Normalized correlation between different data streams.}
  \label{fig_correlation}
\end{figure*}

Fig. \ref{fig_distance} demonstrates the decreasing trend in achievable rate as the distance between the Tx- and Rx-CAPAs increases. This behavior can be attributed to two main factors: the increased free-space path loss and the reduction in spatial DoFs in LoS MIMO channels \cite{10220205}. Additionally, there is a substantial performance gain achieved by multiple data streams than single data streams (i.e., $N = 1$) when the Tx- and Rx-CAPAs are close due to the enhanced spatial DoFs, which highlights the importance of fully exploiting spatial DoFs in LoS CAPA-MIMO systems. Furthermore, Fig.~\ref{fig_frequency} shows that the achievable rate increases with carrier frequency. This improvement is attributed to the higher spatial DoFs available at higher frequency bands \cite{bjornson2024enabling}. \textcolor{black}{Note that in Fig.~\ref{fig_frequency}, frequency-dependent factors such as free-space path loss and antenna efficiency are  inherently accounted for in the Green's function. This is because the Green's function captures the fundamental EM behavior, from which free-space path loss and antenna efficiency naturally arise.}

\subsection{\textcolor{black}{Evaluation on Discrete Metasurfaces}}
\textcolor{black}{In practice, achieving full analog amplitude and phase control over a truly continuous aperture is challenging. As discussed earlier, metasurface antennas are a common technique used to approximate such a continuous aperture~\cite{shlezinger2021dynamic, 10480441, 10908611, castellanos2025embracing, 10981846}. A metasurface is comprised of a large array of closely spaced, sub-wavelength radiating elements. Using techniques like the Huygens' metasurface \cite{10981846}, it is possible to independently control the analog amplitude and phase of the transmitted wave at each individual element. Although metasurfaces are discrete, their vast number of elements makes conventional spatial-domain SVD beamforming methods highly complex. To address this issue, both a Fourier-based method and the proposed WMMSE approach can be applied. In particular, the Fourier-based approach can be directly extended to discrete metasurfaces, resulting in lower-dimensional channels in the wavenumber domain, as detailed in \cite{9724113} and \cite{10639537}. For the proposed WMMSE method, a discrete beamformer can be derived by sampling the optimized continuous beamformer $\mathbf{w}(\mathbf{s})$ at the specific locations of the metasurface elements.}

\textcolor{black}{In Fig. \ref{fig_metasurface}, we evaluate and compare the performance of the proposed WMMSE approach and the Fourier-based approach using discrete metasurfaces with varying element spacing, noting that smaller spacing corresponds to more elements within a fixed aperture. Each metasurface element is assumed to have a physical area of $0.05 \lambda \times 0.05 \lambda$ \cite{5996700} and is approximated as a continuous current sheet. This assumption leads to an equivalent effective aperture for each element matching its physical dimensions\cite{balanis2016antenna, 10981846}. All other system parameters follow the settings introduced at the beginning of this section. It can be observed from Fig. \ref{fig_metasurface} that when multiple data streams are considered, the proposed WMMSE approach outperforms the Fourier-based approach under small element spacing, i.e., when the metasurface closely approximates a continuous structure. However, as the element spacing increases, the performance of the proposed WMMSE algorithm deteriorates due to the significant mismatch between the optimized continuous beamformer and the practical discrete beamformer. Additionally, when only a single data stream is considered, the proposed WMMSE approach and the Fourier-based approach yield nearly identical performance.}

\subsection{Correlation Between Different Data Streams}

An important feature of point-to-point MIMO channels is that they can be converted into parallel, non-interfering single-input single-output (SISO) channels, allowing each data stream to be independently decoded without SIC while still achieving the channel capacity. Such an ideal decomposition is precisely realized when employing eigendecomposition \eqref{eigendecomposition_K} for the beamforming design \cite{miller2000communicating}. Results shown in Fig.~\ref{fig_correlation}, where we set $A_{\mathrm{T}} = A_{\mathrm{R}} = 0.5 \text{ m}^2$, indicate that the proposed WMMSE algorithm closely achieves this decomposition. Specifically, the correlation between the $n$-th and $m$-th data streams is defined as 
\begin{equation}
  \xi_{n,m} = \left|\int_{\mathcal{S}_{\mathrm{R}}} \int_{\mathcal{S}_{\mathrm{T}}} v_n^H(\mathbf{r}) h(\mathbf{r}, \mathbf{s}) w_m(\mathbf{s}) d \mathbf{s} d \mathbf{r}\right|^2,
\end{equation}
which can be interpreted as the desired signal power for $m = n$ or the interference caused by the $m$-th data stream when decoding the $n$-th data stream for $m \neq n$. More particularly, the beamformer $w_m(\mathbf{s})$ is obtained using the proposed WMMSE algorithm, and the receiver $v_n(\mathbf{r})$ is designed as the optimal MMSE receiver in \eqref{MMSE_receiver_no_SIC} without SIC. Fig. \ref{fig_correlation} illustrates that near-perfect orthogonality between data streams is achievable for $N = 6$ data streams. While for $N = 8$ and $N=10$, the orthogonality is slightly reduced, but the inter-stream interference is still maintained at a low level. Furthermore, when $N = 10$, it can be observed that the strength of the signal is high for only the first several data streams. This is because of the limited spatial DoFs of the LoS channel.  Finally, considering the marginal achievable rate improvement from $42.08$ bit/s/Hz to $42.26$ bit/s/Hz when increasing from $6$ to $10$ data streams, it would be more efficient in practice to avoid transmitting a large number of data streams, which helps to reduce receiver complexity while maintaining high performance.

\section{Conclusions} \label{sec:conclusion}

This paper has proposed an efficient WMMSE algorithm to address beamforming optimization in point-to-point CAPA-MIMO systems, effectively overcoming the discretization loss and high computational complexity associated with conventional Fourier-based approaches. The proposed algorithm also shows promise for extension to multi-user CAPA-MIMO beamforming optimization, which is a compelling direction for future research. \textcolor{black}{Additionally, channel estimation remains a critical issue for fully harnessing the CAPA beamforming designs. A promising direction to address the continuous channel estimation challenge in CAPA involves compressive sensing and parametric estimation techniques, which exploit the intrinsic structure of the channel. Further research on these methods is essential to enhance practical system performance.}

\begin{appendices}

    \begin{figure*}[b]
        \vspace*{8pt}
        \hrulefill
        \vspace*{8pt}
        \begin{align} \label{SIC_rate_reformulation}
            R_n = &\log \left( 1 + \frac{1}{\sigma^2} \int_{\mathcal{S}_{\mathrm{R}}}  \left| e_n(\mathbf{r}) \right|^2 d \mathbf{r}  - \frac{1}{\sigma^4} \mathbf{q}_n^H \left( \mathbf{I}_{N-n} + \frac{1}{\sigma^2} \mathbf{Q}_n \right)^{-1} \mathbf{q}_n \right) \nonumber \\
            \overset{(a)}{=} & \log  \det \left( \everymath{\displaystyle} \begin{bmatrix} 
                \left(1 + \frac{1}{\sigma^2} \int_{\mathcal{S}_{\mathrm{R}}}  \left| e_n(\mathbf{r}) \right|^2 d \mathbf{r}\right) & \frac{1}{\sigma^2}\mathbf{q}_n^H \\
                \frac{1}{\sigma^2}\mathbf{q}_n & \left(\mathbf{I}_{N-n} + \frac{1}{\sigma^2} \mathbf{Q}_n\right)
            \end{bmatrix} \right)  - \log \det \left( \mathbf{I}_{N-n} + \frac{1}{\sigma^2} \mathbf{Q}_n \right) \nonumber \\
            = & \log \det \left( \mathbf{I}_{N-n+1} + \frac{1}{\sigma^2} \mathbf{Q}_{n-1} \right) - \log \det \left( \mathbf{I}_{N-n} + \frac{1}{\sigma^2} \mathbf{Q}_n \right). \tag{92}
        \end{align}
    \end{figure*}

    \section{Proof of Theorem \ref{theorem_SU_capacity}} \label{theorem_SU_capacity_proof}

    In this appendix, we derive the achievable rate based on the MMSE-SIC receiver, which is shown to be optimal for MIMO systems \cite{tse2005fundamentals}. Without loss of generality, we assume that the SIC is sequentially carried out from the first to the $N$-th data stream. Under this assumption, the signal for decoding the $n$-th data stream after the SIC is given by 
    \begin{align} \label{SIC_signal}
        y_n(\mathbf{r}) = &\int_{\mathcal{S}_{\mathrm{T}}}  h(\mathbf{r}, \mathbf{s}) w_n (\mathbf{s}) c_n  d \mathbf{s} \nonumber \\
        &+ \sum_{j > n} \int_{\mathcal{S}_{\mathrm{T}}} h(\mathbf{r}, \mathbf{s}) w_j (\mathbf{s}) c_j  d \mathbf{s} + z(\mathbf{r}).
    \end{align}
    Applying a receiver $v_n(\mathbf{r})$ to $y_n(\mathbf{r})$ yields 
    \begin{equation} \label{SIC_signal_receiver}
        \widetilde{y}_n = \int_{\mathcal{S}_{\mathrm{R}}} v_n^H(\mathbf{r}) e_n(\mathbf{r}) c_n d \mathbf{r} + \sum_{j > n} \int_{\mathcal{S}_{\mathrm{R}}}  v_n^H(\mathbf{r}) e_j(\mathbf{r}) c_j  d \mathbf{r} + z_n,
    \end{equation}  
    where 
    \begin{equation}
        e_n(\mathbf{r}) =  \int_{\mathcal{S}_{\mathrm{T}}} h(\mathbf{r}, \mathbf{s}) w_n(\mathbf{s}) d \mathbf{s}, \quad z_n = \int_{\mathcal{S}_{\mathrm{R}}} v_n^H(\mathbf{r}) z(\mathbf{r}) d \mathbf{r}.
    \end{equation}
    More particularly, $z_n$ is the effective noise after the receiver. According to \cite[Lemma 1]{ouyang2024performance}, it follows a distribution of
    \begin{equation}
        z_n \sim \mathcal{CN} \left(0, \sigma^2 \int_{\mathcal{S}_{\mathrm{R}}} |v_n(\mathbf{r})|^2 d \mathbf{r} \right).
    \end{equation}
    Consequently, the SINR for decoding $c_n$ is given by 
    \begin{equation} \label{SIC_SINR}
        \gamma_n = \frac{\left| \int_{\mathcal{S}_{\mathrm{R}}} v_n^H(\mathbf{r}) e_n(\mathbf{r}) d \mathbf{r} \right|^2 }{\sum_{j > n} \left| \int_{\mathcal{S}_{\mathrm{R}}} v_n^H(\mathbf{r}) e_j(\mathbf{r}) d \mathbf{r} \right|^2 + \sigma^2 \int_{\mathcal{S}_{\mathrm{R}}} |v_n(\mathbf{r})|^2 d \mathbf{r} }.
    \end{equation}
    Following \cite[Theorem 3]{ouyang2024performance}, the optimal $v_n(\mathbf{r})$ for maximizing $\gamma_n$ is the MMSE receiver, which is given by 
    \begin{align} \label{MMSE-SIC_receiver}
        v_n^{\mathrm{MMSE}}(\mathbf{r}) = e_n(\mathbf{r}) - \frac{1}{\sigma^2}\mathbf{e}_n(\mathbf{r}) \left( \mathbf{I}_{N-n} + \frac{1}{\sigma^2}\mathbf{Q}_n \right)^{-1} \mathbf{q}_n,
    \end{align} 
    where 
    \begin{align} \label{SIC_Q_matrix}
        \mathbf{Q}_n = &\int_{\mathcal{S}_{\mathrm{R}}} \mathbf{e}_n^H(\mathbf{r}) \mathbf{e}_n(\mathbf{r}) d \mathbf{r} \in \mathbb{C}^{(N-n) \times (N-n)}, \\
        \mathbf{q}_n = &\int_{\mathcal{S}_{\mathrm{R}}} \mathbf{e}_n^H(\mathbf{r}) e_n(\mathbf{r}) d \mathbf{r} \in \mathbb{C}^{(N-n) \times 1},
    \end{align}
    and $\mathbf{e}_n(\mathbf{r}) = [e_{n+1}(\mathbf{r}),\dots,e_N(\mathbf{r})] \in \mathbb{C}^{1 \times (N-n)}$. By substituting \eqref{MMSE-SIC_receiver} into \eqref{SIC_SINR}, we obtain the following optimal SINR (see Section V-C of \cite{ouyang2024performance} for detailed derivations):
    \begin{align}
        \gamma_n^{\mathrm{MMSE}} = \frac{1}{\sigma^2} \int_{\mathcal{S}_{\mathrm{R}}} & \left| e_n(\mathbf{r}) \right|^2 d \mathbf{r} \nonumber \\
        & - \frac{1}{\sigma^4} \mathbf{q}_n^H \left( \mathbf{I}_{N-n} + \frac{1}{\sigma^2} \mathbf{Q}_n \right)^{-1} \mathbf{q}_n.
    \end{align}
    Therefore, the achievable rate realized by the MMSE-SIC receiver is given by
    \begin{align}
        R = \sum_{n=1}^N \log \left(1 + \gamma_n^{\mathrm{MMSE}}\right).
    \end{align}
    However, the above expression is non-tractable due to the complicated form of $\gamma_n^{\mathrm{MMSE}}$. To simplify it, we reformulate the achievable rate of each data stream, i.e., $R_n = \log(1 + \gamma_n^{\mathrm{MMSE}})$, as given in \eqref{SIC_rate_reformulation}. In particular, step $(a)$ in \eqref{SIC_rate_reformulation} follows the determinant of block matrices, which states that \cite{silvester2000determinants} 
    \setcounter{equation}{92}
    \begin{equation}
        \det \left( \begin{bmatrix}
            \mathbf{A} & \mathbf{B} \\
            \mathbf{C} & \mathbf{D}
        \end{bmatrix} \right) = \det \left( \mathbf{A} - \mathbf{B} \mathbf{D}^{-1} \mathbf{C} \right) \det \left(\mathbf{D}\right).
    \end{equation}
    Based on the results in \eqref{SIC_rate_reformulation}, the achievable rate $R$ can be reformulated as
    \begin{align} \label{SIC_sum}
        R &= \sum_{n=1}^{N-1} \Bigg( \log \det \left( \mathbf{I}_{N-n+1} + \frac{1}{\sigma^2} \mathbf{Q}_{n-1} \right) \nonumber \\
        & - \log \det \left( \mathbf{I}_{N-n} + \frac{1}{\sigma^2} \mathbf{Q}_n \right) \Bigg) 
        = \log \det \left( \mathbf{I}_{N} + \frac{1}{\sigma^2} \mathbf{Q}_0 \right). 
    \end{align}
    By defining $\mathbf{Q} \triangleq \mathbf{Q}_0$, the expression of the achievable rate in \eqref{SU_capacity} can be obtained. The proof is thus completed.

    \section{Proof of Theorem \ref{theorem_2}} \label{theorem_2_proof}

    This theorem can be proven by following the approach used in the proof for conventional discrete arrays \cite{shi2011iteratively}. First, it can be shown that when other optimization variables are fixed, the optimal $\mathbf{v}(\mathbf{r})$ corresponds to the MMSE receiver given in \eqref{MMSE_receiver_no_SIC}, while the optimal $\mathbf{U}$ is given by $\mathbf{E}^{-1}$. Substituting these optimal solutions into problem \eqref{MSE_minimization_problem} transforms it into 
    \begin{equation}
        \min_{\mathbf{w}(\mathbf{s})} \quad N - \log \det (\mathbf{E}_{\mathrm{MMSE}}^{-1}),
    \end{equation} 
    which, according to \eqref{MMSE_MSE_matrix}, is equivalent to problem \eqref{unconstrained_problem}. The proof is thus completed.

    \section{Proof of Proposition \ref{optimal_w_s}} \label{optimal_w_s_proof}

    To simplify the derivation of the optimal $\mathbf{w}(\mathbf{s})$, we first define the following notations: 
    \begin{equation}
        \mathbf{g}(\mathbf{s}) = \int_{\mathcal{S}_{\mathrm{R}}} h^H(\mathbf{r}, \mathbf{s}) \mathbf{v}(\mathbf{r}) d \mathbf{r},  \quad \mathbf{V} = \int_{\mathcal{S}_{\mathrm{R}}} \mathbf{v}^H(\mathbf{r}) \mathbf{v}(\mathbf{r})  d \mathbf{r}.
    \end{equation}
    Then, the MSE matrix in \eqref{MSE_matrix} can be reformulated as 
    \begin{align}
        \mathbf{E} 
        = & \left( \mathbf{I}_N - \int_{\mathcal{S}_{\mathrm{T}}} \!\! \mathbf{g}^H(\mathbf{s}) \mathbf{w}(\mathbf{s}) d \mathbf{s} \right) \left( \mathbf{I}_N - \int_{\mathcal{S}_{\mathrm{T}}} \!\! \mathbf{g}^H(\mathbf{s}) \mathbf{w}(\mathbf{s}) d \mathbf{s} \right)^H \nonumber \\
        &\hspace{3.5cm} + \frac{\sigma^2}{P_{\mathrm{T}}} \mathbf{V} \int_{\mathcal{S}_{\mathrm{T}}} \|\mathbf{w} (\mathbf{s})\|^2 d \mathbf{s}.
    \end{align}
    When all optimization parameters are fixed except for $\mathbf{w}(\mathbf{s})$, problem \eqref{MSE_minimization_problem} reduces to
    \begin{equation}
        \min_{\mathbf{w}(\mathbf{s})} \quad G\big(\mathbf{w}(\mathbf{s})\big) = \mathrm{Tr}(\mathbf{U} \mathbf{E}).
    \end{equation}
    This problem is a functional optimization problem, which can be solved using the CoV method. Specifically, the optimal $\mathbf{w}(\mathbf{s})$ is determined by satisfying the following condition:
    \begin{equation} \label{CoV_condition}
        \left. \frac{d G \big(\mathbf{w}(\mathbf{s}) + \epsilon \boldsymbol{\eta}(\mathbf{s}) \big) }{d \epsilon} \right|_{\epsilon = 0} = 0,
    \end{equation} 
    where $\boldsymbol{\eta}(\mathbf{s}) \in \mathbb{C}^{1 \times N}$ is any arbitrary smooth function with each entry defined on $\mathcal{S}_{\mathrm{T}}$. To leverage the above condition, we first give the explicit expression for $\Phi(\epsilon) \triangleq G \big(\mathbf{w}(\mathbf{s}) + \epsilon \boldsymbol{\eta}(\mathbf{s}) \big)$, given by:
    \begin{align} \label{Phi_expression}
        \Phi(\epsilon) = 2 \epsilon \Re \left\{ \int_{\mathcal{S}_{\mathrm{T}}} \mathbf{p} (\mathbf{s})  \boldsymbol{\eta}^H(\mathbf{s}) d \mathbf{s} \right\} + \Psi(\epsilon^2) + C,
    \end{align}
    where $\Psi(\epsilon^2)$ collect terms related to $\epsilon^2$, $C$ is a constant, and    
    \begin{align}
        \mathbf{p}(\mathbf{s}) = \mathbf{g}(\mathbf{s}) \mathbf{U} \left( \int_{\mathcal{S}_{\mathrm{T}}} \mathbf{g}^H(\mathbf{z}) \mathbf{w}(\mathbf{z}) d \mathbf{z} - \mathbf{I}_N \right) \nonumber \\
        + \frac{\sigma^2  \mathrm{Tr} \left( \mathbf{U} \mathbf{V} \right)}{P_{\mathrm{T}}} \mathbf{w}(\mathbf{s}). 
    \end{align}
    Substituting \eqref{Phi_expression} into \eqref{CoV_condition} yields
    \begin{equation}
        \Re \left\{ \int_{\mathcal{S}_{\mathrm{T}}} \mathbf{p} (\mathbf{s})  \boldsymbol{\eta}^H(\mathbf{s}) d \mathbf{s} \right\} = 0.
    \end{equation}
    Following the fundamental lemma of CoV given in \cite[Lemma 1]{10938678}, it must hold that $\mathbf{p} (\mathbf{s}) = \mathbf{0}$ to satisfy the above conditions for any arbitrary $\boldsymbol{\eta}(\mathbf{s})$, leading to the following result:
    \begin{equation} \label{F_equation}
        \mathbf{w}(\mathbf{s}) = \varepsilon \mathbf{g}(\mathbf{s}) \mathbf{U} \left( \mathbf{I}_N - \int_{\mathcal{S}_{\mathrm{T}}} \mathbf{g}^H(\mathbf{z}) \mathbf{w}(\mathbf{z}) d \mathbf{z} \right),
    \end{equation}  
    where $\varepsilon = \frac{P_{\mathrm{T}}}{\sigma^2 \mathrm{Tr}(\mathbf{U} \mathbf{V})}$. 
    However, obtaining an explicit expression for $\mathbf{w}(\mathbf{s})$ from \eqref{F_equation} remains challenging, as $\mathbf{w}(\mathbf{s})$ appears on both sides of the equation. To solve it, we multiply both sides of \eqref{F_equation} by $\mathbf{g}^H(\mathbf{s})$ and integrate over $d \mathbf{s}$, resulting in 
    \begin{equation} \label{Matrix_form_optimal_condition}
        \mathbf{\Lambda} = \varepsilon \mathbf{G} \mathbf{U} \left( \mathbf{I}_N - \mathbf{\Lambda}\right),
    \end{equation} 
    where 
    \begin{equation}
        \mathbf{\Lambda} = \int_{\mathcal{S}_{\mathrm{T}}} \mathbf{g}^H(\mathbf{s}) \mathbf{w}(\mathbf{s}) d \mathbf{s}, \quad \mathbf{G} = \int_{\mathcal{S}_{\mathrm{T}}} \mathbf{g}^H(\mathbf{s}) \mathbf{g}(\mathbf{s}) d \mathbf{s}.
    \end{equation}
    Following \eqref{Matrix_form_optimal_condition}, matrix $\mathbf{\Lambda}$ can be calculated as 
    \begin{equation} \label{optimal_Lambda}
        \mathbf{\Lambda} = \left( \frac{1}{\varepsilon} \mathbf{I}_N +   \mathbf{G} \mathbf{U} \right)^{-1} \mathbf{G} \mathbf{U}.
    \end{equation} 
    By substituting \eqref{optimal_Lambda} into \eqref{F_equation}, the optimal $\mathbf{w}(\mathbf{s})$ can be obtained as  
    \begin{align} \label{appendix_c_optimal_solution}
        \mathbf{w}(\mathbf{s}) = &\varepsilon \mathbf{g}(\mathbf{s}) \mathbf{U} \left( \mathbf{I}_N - \left( \frac{1}{\varepsilon} \mathbf{I}_N +  \mathbf{G} \mathbf{U} \right)^{-1} \hspace{-0.2cm} \mathbf{G} \mathbf{U} \right) \nonumber \\
        = & \mathbf{g}(\mathbf{s}) \mathbf{U}  \left(\frac{1}{\varepsilon} \mathbf{I}_N + \mathbf{G} \mathbf{U} \right)^{-1}.
    \end{align}

    \section{Extension of WMMSE for \\ Correlated Power Constraints} \label{appendix_extension}

    In this appendix, we discuss the extension of the WMMSE algorithm for solving the more general correlated power constraint given below:
    \begin{equation} \label{power_constraint_correlated}
        \int_{\mathcal{S}_{\mathrm{T}}} \int_{\mathcal{S}_{\mathrm{T}}} \mathbf{\mathbf{w}}(\mathbf{s}) c_{\mathrm{T}} (\mathbf{s}-\mathbf{z}) \mathbf{w}^H(\mathbf{z}) d \mathbf{z} d \mathbf{s} \le P_{\mathrm{T}},
    \end{equation}
    where $c_{\mathrm{T}} (\mathbf{s}-\mathbf{z})$ characterize the correlation between points $\mathbf{s}$ and $\mathbf{z}$ at the Tx-CAPA. One of the primary reasons for correlation is mutual coupling, which refers to the electromagnetic interaction between individual radiating elements when placed in close proximity \cite{pizzo2025mutual}. When $c_{\mathrm{T}} (\mathbf{s}-\mathbf{z}) = \delta(\mathbf{s}-\mathbf{z})$, i.e., there is no correlation, the power constraint \eqref{power_constraint_correlated} reduces to \eqref{power_constraint}. Similar to \textbf{Lemma \ref{equal_power_lemma}}, maximizing the achievable rate subject to the new power constraint \eqref{power_constraint_correlated} can be formulated as an unconstrained problem as  
    \begin{equation} \label{unconstrained_problem_MC}
        \max_{\mathbf{w}(\mathbf{s})} \quad \widetilde{R} = \log \det \left( \mathbf{I}_N + \frac{1}{\overline{\widetilde{\sigma}}^2} \mathbf{Q} \right),
    \end{equation}
    where 
    \begin{equation}
        \label{expression_sigma_MC}
        \overline{\widetilde{\sigma}}^2 = \frac{\sigma^2}{P_{\mathrm{T}}} \int_{\mathcal{S}_{\mathrm{T}}} \int_{\mathcal{S}_{\mathrm{T}}} \mathbf{\mathbf{w}}(\mathbf{s}) c_{\mathrm{T}} (\mathbf{s}-\mathbf{z}) \mathbf{w}^H(\mathbf{z}) d \mathbf{z} d \mathbf{s}.
    \end{equation}
    Following \textbf{Theorem \ref{theorem_2}}, problem \eqref{unconstrained_problem_MC} can be transformed into the following equivalent MSE minimization problem:
    \begin{equation} \label{MSE_minimization_problem_MC}
        \min_{\mathbf{w}(\mathbf{s}), \mathbf{v}(\mathbf{r}), \mathbf{U}} \quad \mathrm{Tr} \left( \mathbf{U} \widetilde{\mathbf{E}} \right) - \log \det (\mathbf{U}),
    \end{equation}
    where $\widetilde{\mathbf{E}}$ is given by 
    \begin{align}
        \widetilde{\mathbf{E}} 
        = & \left( \mathbf{I}_N - \int_{\mathcal{S}_{\mathrm{T}}} \!\! \mathbf{g}^H(\mathbf{s}) \mathbf{w}(\mathbf{s}) d \mathbf{s} \right) \left( \mathbf{I}_N - \int_{\mathcal{S}_{\mathrm{T}}} \!\! \mathbf{g}^H(\mathbf{s}) \mathbf{w}(\mathbf{s}) d \mathbf{s} \right)^H \nonumber \\
        &\hspace{0.5cm} + \frac{\sigma^2}{P_{\mathrm{T}}} \mathbf{V} \int_{\mathcal{S}_{\mathrm{T}}} \int_{\mathcal{S}_{\mathrm{T}}} \mathbf{\mathbf{w}}(\mathbf{s}) c_{\mathrm{T}} (\mathbf{s}-\mathbf{z}) \mathbf{w}^H(\mathbf{z}) d \mathbf{z} d \mathbf{s}.
    \end{align}
    
    Following the procedure in Section \ref{sec:WMMSE}, the optimal solutions of $\mathbf{v}(\mathbf{r})$ and $\mathbf{U}$ to problem \eqref{MSE_minimization_problem_MC} can be derived as 
    \begin{align} 
        \label{MC_v}
        \mathbf{v}^{\star}(\mathbf{r}) &= \widetilde{\mathbf{v}}_{\mathrm{MMSE}} = \frac{1}{\overline{\widetilde{\sigma}}^2} \mathbf{e}(\mathbf{r}) \left( \mathbf{I}_N + \frac{1}{\overline{\widetilde{\sigma}}^2}  \mathbf{Q} \right)^{-1}, \\
        \label{MC_U}
        \mathbf{U}^{\star} &= \widetilde{\mathbf{E}}_{\mathrm{MMSE}}^{-1} = \mathbf{I}_N + \frac{1}{\overline{\widetilde{\sigma}}^2} \mathbf{Q}.
    \end{align} 

    The optimal solution of $\mathbf{w}(\mathbf{s})$ can be obtained by minimizing $\widetilde{G}(\mathbf{w}(\mathbf{s})) = \mathrm{Tr}(\mathbf{U} \widetilde{\mathbf{E}})$ using the CoV method as in Appendix \ref{optimal_w_s_proof}. In particular, define $\widetilde{\Phi}(\epsilon) = \widetilde{G}(\mathbf{w}(\mathbf{s}) + \epsilon \boldsymbol{\eta}(\mathbf{s}))$, whose explicit expression is given by 
    \begin{align} \label{Phi_expression_MC}
        \widetilde{\Phi}(\epsilon) = 2 \epsilon \Re \left\{ \int_{\mathcal{S}_{\mathrm{T}}} \widetilde{\mathbf{p}} (\mathbf{s})  \boldsymbol{\eta}^H(\mathbf{s}) d \mathbf{s} \right\} + \widetilde{\Psi}(\epsilon^2) + \widetilde{C},
    \end{align}
    where $\widetilde{\Psi}(\epsilon^2)$ collect terms related to $\widetilde{\epsilon}^2$, $\widetilde{C}$ is a constant, and    
    \begin{align}
        \widetilde{\mathbf{p}}(\mathbf{s}) = \mathbf{g}(\mathbf{s}) \mathbf{U} \left( \int_{\mathcal{S}_{\mathrm{T}}} \mathbf{g}^H(\mathbf{z}) \mathbf{w}(\mathbf{z}) d \mathbf{z} - \mathbf{I}_N \right) \nonumber \\
        + \frac{1}{\varepsilon} \int_{\mathcal{S}_{\mathrm{T}}} c_{\mathrm{T}}(\mathbf{s}-\mathbf{z}) \mathbf{w}(\mathbf{z}) d \mathbf{z}.
    \end{align}
    Following the fundamental lemma of CoV, it must hold that $\widetilde{\mathbf{p}}(\mathbf{s}) = \mathbf{0}$, which yields
    \begin{align} \label{MC_optimal_condition}
        \int_{\mathcal{S}_{\mathrm{T}}} c_{\mathrm{T}}(\mathbf{s}-\mathbf{z}) \mathbf{w}(\mathbf{z}) d \mathbf{z} = \varepsilon \mathbf{g}(\mathbf{s}) \mathbf{U} \left(\mathbf{I}_N - \int_{\mathcal{S}_{\mathrm{T}}} \mathbf{g}^H(\mathbf{z}) \mathbf{w}(\mathbf{z}) d \mathbf{z} \right).
    \end{align}
    Define $c^{\mathrm{inv}}_{\mathrm{T}}(\mathbf{r}-\mathbf{s})$ such that 
    \begin{equation}
        \int_{\mathcal{S}_{\mathrm{T}}} c^{\mathrm{inv}}_{\mathrm{T}}(\mathbf{r}-\mathbf{s}) c_{\mathrm{T}}(\mathbf{s}-\mathbf{z}) d \mathbf{s} = \delta(\mathbf{r}-\mathbf{z}).
    \end{equation}
    Multiplying both sides of \eqref{MC_optimal_condition} by $c^{\mathrm{inv}}_{\mathrm{T}}(\mathbf{r}-\mathbf{s})$ and integrating over $d \mathbf{s}$ yields
    \begin{align} \label{MC_optimal_condition_2}
        \mathbf{w}(\mathbf{r}) = \varepsilon \widetilde{\mathbf{g}}(\mathbf{r}) \mathbf{U} \left(\mathbf{I}_N - \int_{\mathcal{S}_{\mathrm{T}}}  \mathbf{g}^H(\mathbf{z}) \mathbf{w}(\mathbf{z}) d \mathbf{z}\right),
    \end{align} 
    where 
    \begin{align}
        \widetilde{\mathbf{g}}(\mathbf{r}) = \int_{\mathcal{S}_{\mathrm{T}}}  c^{\mathrm{inv}}_{\mathrm{T}}(\mathbf{r}-\mathbf{s}) \mathbf{g}(\mathbf{s}) d \mathbf{s}.
    \end{align}
    Then, following the same procedure as in \eqref{Matrix_form_optimal_condition}-\eqref{appendix_c_optimal_solution}, the optimal $\mathbf{w}(\mathbf{r})$ can be obtained as 
    \begin{equation}
        \mathbf{w}(\mathbf{r}) = \widetilde{\mathbf{g}}(\mathbf{r}) \mathbf{U}  \left(\frac{1}{\varepsilon} \mathbf{I}_N + \widetilde{\mathbf{G}} \mathbf{U} \right)^{-1},
    \end{equation} 
    where 
    \begin{align}
        \widetilde{\mathbf{G}} = &\int_{\mathcal{S}_{\mathrm{T}}} \mathbf{g}^H(\mathbf{r}) \widetilde{\mathbf{g}}(\mathbf{r}) d \mathbf{r} \nonumber \\
        = &\int_{\mathcal{S}_{\mathrm{T}}} \int_{\mathcal{S}_{\mathrm{T}}} \mathbf{g}^H(\mathbf{r})  c^{\mathrm{inv}}_{\mathrm{T}}(\mathbf{r}-\mathbf{s}) \mathbf{g}(\mathbf{s}) d \mathbf{s} d \mathbf{r}.
    \end{align}

    Based on the results in \eqref{MC_v}, \eqref{MC_U}, and \eqref{MC_optimal_condition_2}, the WMMSE algorithm for correlated power constraints can be established. However, obtaining a closed-form expression for $c^{\mathrm{inv}}_{\mathrm{T}}(\mathbf{r}-\mathbf{s})$ remains generally challenging unless the term assumes a special form, as demonstrated in \cite[Lemma 2]{10938678}. In general cases, no systematic method exists to explicitly compute $c^{\mathrm{inv}}_{\mathrm{T}}(\mathbf{r}-\mathbf{s})$, necessitating a scenario-specific analysis. This limitation highlights an open problem and will be addressed in future work.
\end{appendices}

\balance
\bibliographystyle{IEEEtran}
\bibliography{reference/mybib}

\end{document}